\documentclass[acmsmall,authorversion,screen,nonacm]{acmart} 

\AtBeginDocument{%
  }

\setcopyright{none} 
\renewcommand\footnotetextcopyrightpermission{} 
\copyrightyear{2026}
\acmYear{2026}
\acmDOI{}
\acmISBN{2026}

\acmConference[SIGGRAPH '26]{ACM SIGGRAPH 2026 Art Papers}{July 19--23, 2026}{Los Angeles, CA, USA}

\acmJournal{PACMCGIT}

\acmISBN{978-1-4503-XXXX-X/26/07}

\usepackage{multirow}
\usepackage[table,xcdraw]{xcolor}
\usepackage{tabularx}
\usepackage{makecell}
\usepackage{graphicx}
\usepackage{subcaption}

\begin{document}
\thanks{© Author 2026. This is the author's version of the work. It is posted here for your personal use. Not for redistribution. The definitive version of record will be published in Proceedings of the ACM on Computer Graphics and Interactive Techniques (PACMCGIT)}

\title[Large-Scale Photogrammetric Documentation of St. John's Co-Cathedral]{Large-Scale Photogrammetric Documentation of St. John's Co-Cathedral: A Workflow for Cultural Heritage Preservation}

\author{Matthew Kenely}
\affiliation{%
  \institution{SeyTravel Ltd}
  \city{Birkirkara}
  \country{Malta}}
\email{matthew.kenely@seytravel.com}

\author{Mark Bugeja}
\affiliation{%
  \institution{University of Malta}
  \city{Msida}
  \country{Malta}}
\email{mark.bugeja@um.edu.mt}

\author{Andre Grima}
\affiliation{%
  \institution{Stargate Studios}
  \city{Mosta}
  \country{Malta}}
\email{andre.grima@stargatestudios.com.mt}

\author{Peter Pullicino}
\affiliation{%
  \institution{Stargate Studios}
  \city{Mosta}
  \country{Malta}}
\email{peter.pullicino@stargatestudios.com.mt}

\author{Matthew Pullicino}
\affiliation{%
  \institution{Stargate Studios}
  \city{Mosta}
  \country{Malta}}
\email{matthew.pullicino@stargatestudios.com.mt}

\author{Dylan Seychell}
\affiliation{%
  \institution{University of Malta}
  \city{Msida}
  \country{Malta}}
\email{dylan.seychell@um.edu.mt}

\renewcommand{\shortauthors}{Kenely et al.}


\begin{abstract}
We present a comprehensive methodology for the large-scale photogrammetric documentation of St. John's Co-Cathedral in Valletta, Malta, a UNESCO World Heritage site renowned for its ornate Baroque architecture and Caravaggio masterpieces. Over seven nights of evening-only data collection, we captured 99,000 images using DSLR cameras, drone photography, and LIDAR scanning to create a highly detailed 3D reconstruction comprising 25-30 billion triangles. This paper documents our complete workflow for cultural heritage preservation, addressing the unique challenges of digitizing complex baroque architectural spaces with highly reflective metallic surfaces, dark materials, intricate tapestries, and restricted access. We detail our pipeline from multi-modal data acquisition through processing, including strategic image grading and AI-assisted denoising to address low-light grain, extensive LIDAR point cloud cleanup, hybrid photogrammetric reconstruction using RealityCapture, and mesh subdivision strategies for real-time visualization engines. Our methodology combines automated workflows with necessary manual intervention to handle the scale and complexity of the project, with particular attention to reflective surface challenges characteristic of baroque heritage sites. We also present preliminary experiments with Gaussian splatting as a complementary representation technique. The resulting digital archive serves multiple preservation purposes including disaster recovery documentation, conservation analysis, virtual tourism, and scholarly research. This work provides a detailed, replicable workflow for heritage professionals undertaking similar large-scale architectural documentation projects, addressing the practical challenges of applying photogrammetric methods in complex real-world heritage scenarios.
\end{abstract}

\begin{teaserfigure}
  \includegraphics[width=\textwidth]{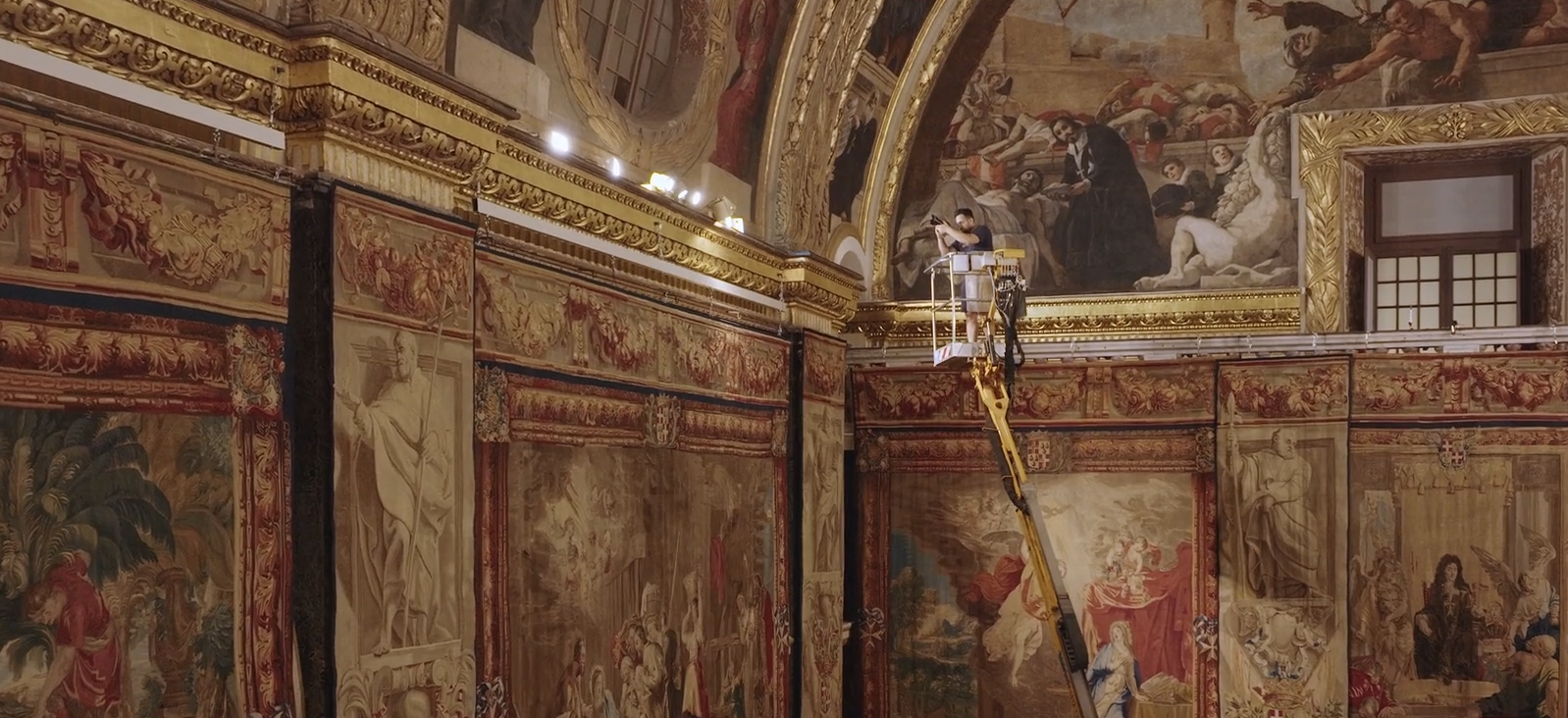}
  \caption{Data acquisition in progress at St. John's Co-Cathedral. A technician on a lift platform captures high-resolution photographs of the ornate Baroque tapestries, gilded architectural details, and painted ceilings. Over seven nights, 99,000 images were collected to create a comprehensive 3D reconstruction of this UNESCO World Heritage site.}
  \Description{Interior view of St. John's Co-Cathedral showing a photographer on a yellow lift platform photographing ornate tapestries and baroque architectural details.}
  \label{fig:teaser}
\end{teaserfigure}



\maketitle

\section{Introduction}

St. John's Co-Cathedral in Valletta, Malta, exemplifies baroque architectural heritage at its finest, featuring elaborate gilt carvings, marble tombstones, and two Caravaggio paintings. As a UNESCO World Heritage site, its preservation demands comprehensive documentation methods beyond traditional photographs and measured drawings \cite{Letellier2007, Remondino2011}.

\begin{figure}[h]
  \centering
  \includegraphics[width=\columnwidth]{figures/Wide-angle_cathedral_interior_panorama.jpg}
  \caption{St. John's Co-Cathedral nave interior showing the ornate baroque ceiling paintings, gilded architectural details, tapestries along the walls, and intricate marble floor inlay work. The cathedral's exceptional baroque decoration and scale present significant challenges for comprehensive digital documentation.}
  \Description{Wide-angle photograph of cathedral nave showing elaborate baroque decoration with painted ceiling vaults, ornate gold leaf architectural elements, wall tapestries, and decorative marble floor}
  \label{fig:cathedral-nave}
\end{figure}


Recent advances in photogrammetry, LIDAR scanning, and 3D reconstruction enable comprehensive digital replicas of heritage sites \cite{Remondino2011, Aicardi2018}. However, baroque spaces present acute challenges: highly reflective metallic surfaces, intricate three-dimensional details, restricted shooting angles, and massive data volumes \cite{DelPozo2017, Munumer2020}.

This paper documents our methodology for large-scale photogrammetric reconstruction of St. John's Co-Cathedral (Figure~\ref{fig:cathedral-nave}). Over seven nights (7pm onwards), we captured 99,000 images using DSLR cameras, drones, and LIDAR scanners, producing a reconstruction exceeding 25 billion triangles. Client priorities emphasizing the tapestry-adorned nave influenced our allocation: 2.5 nights for the nave, 2.5 for chapels, and 2 for the altar.

Our contributions address cultural heritage and computer graphics communities through: a comprehensive, replicable workflow combining DSLR, drone, and LIDAR data; practical solutions to baroque-specific challenges including reflective surfaces and access limitations; methods for processing extremely large datasets (99k images, 25+ billion triangles) for both archival and interactive visualization; and guidelines for multi-night acquisition in active heritage sites. This methodology prioritizes accessibility for heritage professionals through commercial software with targeted custom scripting. Rather than novel algorithms, we document a proven workflow for this scale and complexity, addressing the gap between theoretical photogrammetric methods and practical application in challenging real-world scenarios.

\section{Related Work}

Digital documentation of cultural heritage has evolved from individual artifact scanning to comprehensive architectural digitization. Foundational projects established key methodologies: the Digital Michelangelo Project \cite{Levoy2000} addressed challenges of scale in sculpture scanning, while the Bamiyan Buddhas reconstruction \cite{Gruen2004} demonstrated the critical preservation value of digital records. More recent work has applied Structure from Motion to historical photographs for virtual recovery of lost architectural phases \cite{Condorelli2023} and combined photogrammetry with laser scanning for detailed artifact digitization \cite{Bertocci2024}.

Multi-modal integration combining photogrammetry and LIDAR has become standard practice for heritage recording, leveraging geometric accuracy of laser scanning with textural richness of photography \cite{Remondino2011, DeFino2023}. Digital twin frameworks extending these approaches toward long-term monitoring have also been proposed \cite{Kong2023}. Large-scale ecclesiastical interiors present distinct challenges from smaller-scale recording, requiring strategies for spatial extent, geometric coherence \cite{Fassi2017, Oreni2014}, and problematic materials including reflective gilded surfaces \cite{DelPozo2017, Apollonio2021}. Surface reconstruction methods \cite{Kazhdan2006, Berger2017} provide algorithmic foundations, though managing datasets at our scale (99,000 images, 25+ billion triangles) requires strategic subdivision beyond standard capabilities.

3D Gaussian Splatting \cite{Kerbl2023} has achieved real-time rendering while preserving view-dependent effects including reflections and specular highlights challenging for traditional photogrammetric texturing, consistent with our preliminary experiments. Surface-aligned Gaussian representations \cite{Turki2024} suggest hybrid approaches combining structural meshes with neural rendering fidelity. Virtual and extended reality applications for cultural heritage \cite{Bekele2018, Anderson2010} demonstrate growing interest in interactive experiences, highlighting tensions between visual fidelity and real-time performance that our mesh subdivision strategy addresses.

Standards established by ICOMOS \cite{ICOMOS2017} and the Getty Conservation Institute \cite{Letellier2007} emphasize comprehensive documentation and systematic methodology. The destruction of Notre-Dame de Paris underscored both the value of prior digital documentation \cite{neroulidis2024digital} and the urgency of systematic heritage digitization. Despite these advances, few publications document complete end-to-end workflows at this detail level. Our contribution provides comprehensive documentation of decisions, challenges, and solutions across the entire pipeline.

\section{Project Characteristics and Data Acquisition}

St. John's Co-Cathedral's architectural complexity spans multiple distinct spaces: main nave, two side naves, twelve chapels, altar area, and choir section, each requiring different capture strategies. Interior materials posed significant challenges: highly reflective bronze, gold leaf, and metallic surfaces created specular highlights confusing photogrammetric algorithms; dark wood elements required careful exposure management; intricate tapestries demanded high-resolution capture; marble floors with detailed inlay required comprehensive multi-angle coverage; and ceiling paintings necessitated elevated capture positions. As an active worship site and tourist attraction, acquisition was restricted to evening hours (7pm onwards) across seven nights.

\begin{figure*}[t]
  \centering
  \begin{subfigure}[t]{0.48\textwidth}
    \includegraphics[width=\textwidth]{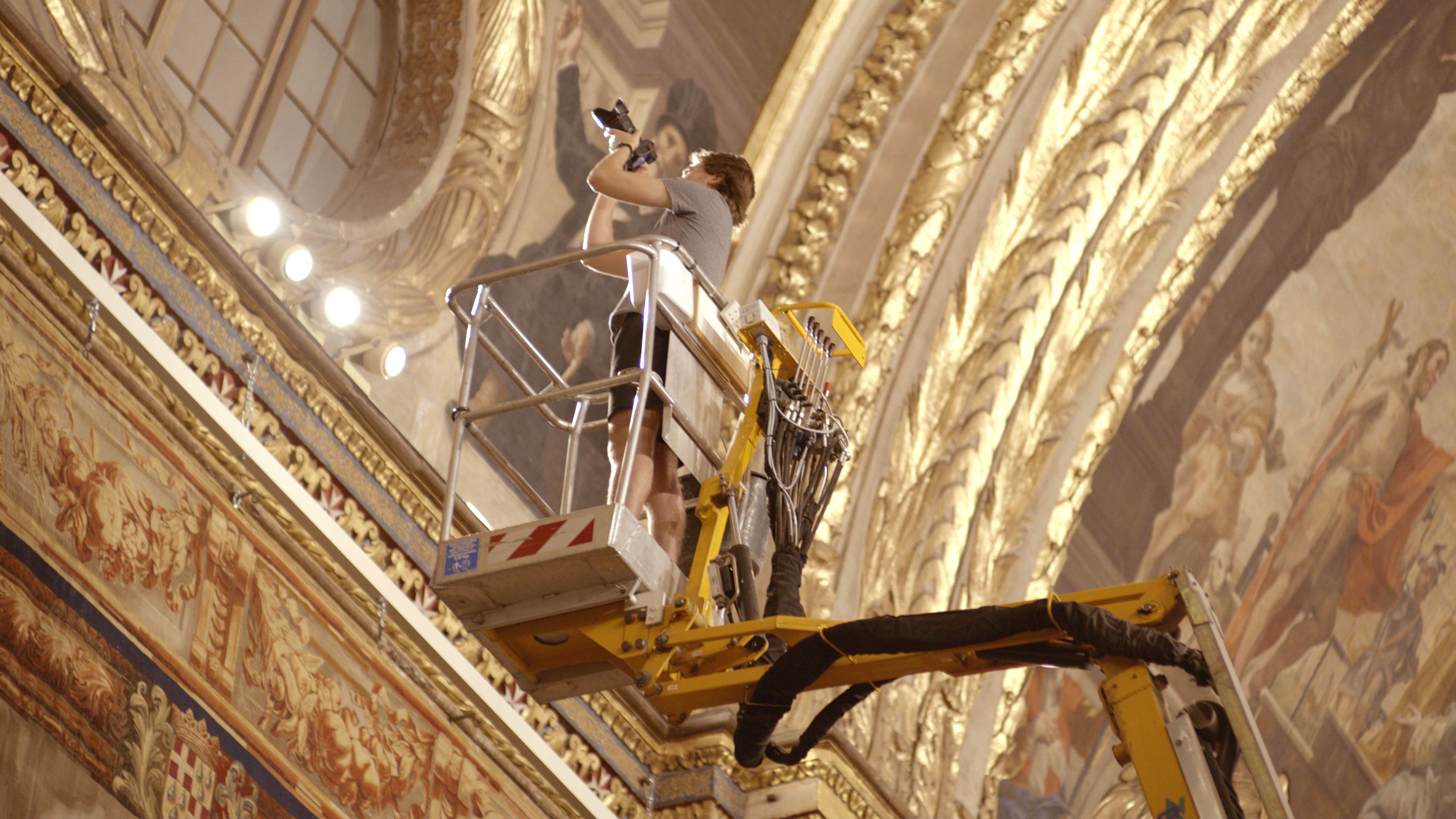}
    \caption{DSLR capture from lift platform}
    \label{fig:acquisition:dslr}
  \end{subfigure}
  \hfill
  \begin{subfigure}[t]{0.48\textwidth}
    \includegraphics[width=\textwidth]{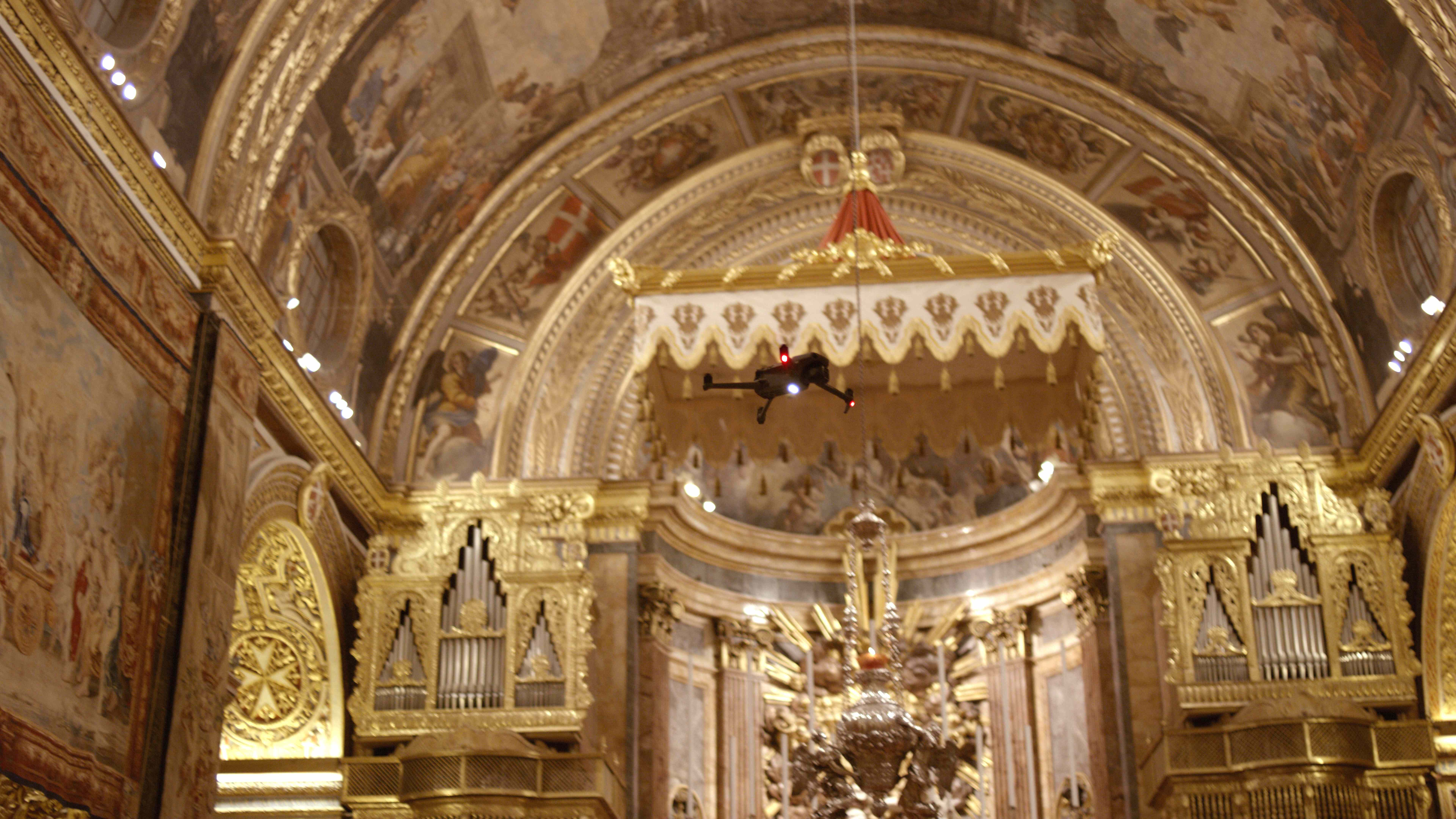}
    \caption{Drone photography of ceiling}
    \label{fig:acquisition:drone}
  \end{subfigure}
  
  \vspace{1em}
  
  \begin{subfigure}[t]{0.48\textwidth}
    \includegraphics[width=\textwidth]{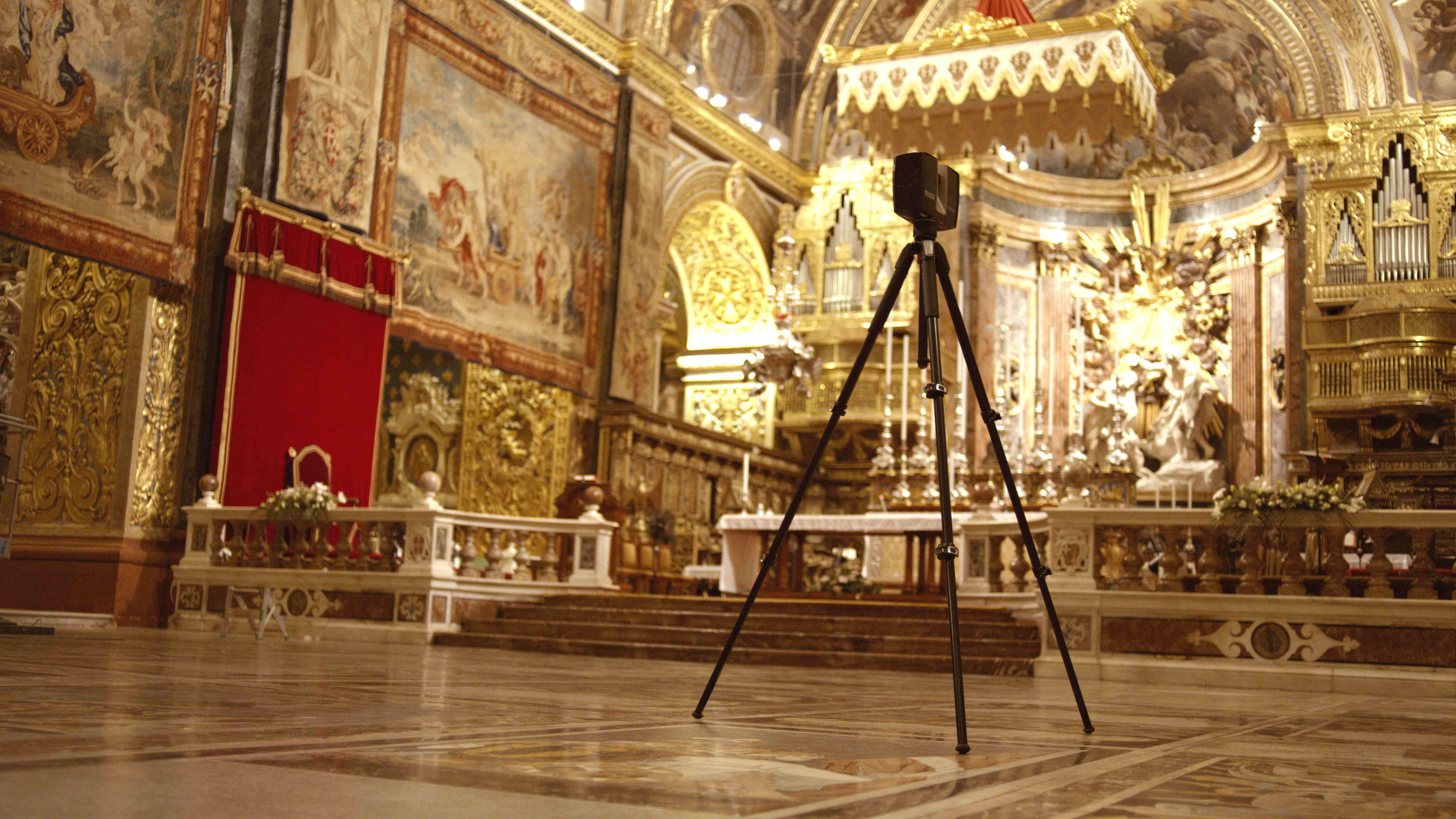}
    \caption{LIDAR scanner positioned in nave}
    \label{fig:acquisition:lidar}
  \end{subfigure}
  \caption{Multi-modal data acquisition approach. (a) A technician on a lift platform captures high-resolution DSLR images of ceiling details and tapestries from elevated positions. (b) Drone photography provides coverage of areas inaccessible to scaffolding, capturing the ornate baroque ceiling paintings at 5K resolution. (c) The Faro LIDAR scanner, positioned centrally, performs 60-minute omnidirectional scans, each generating over 100 million points to provide accurate geometric structure.}
  \Description{Three panels showing different data capture methods: lift platform with photographer, drone in flight, and LIDAR scanner on tripod}
  \label{fig:data-acquisition}
\end{figure*}

\begin{figure*}[t]
  \centering
  \begin{subfigure}[t]{0.48\textwidth}
    \includegraphics[width=\textwidth]{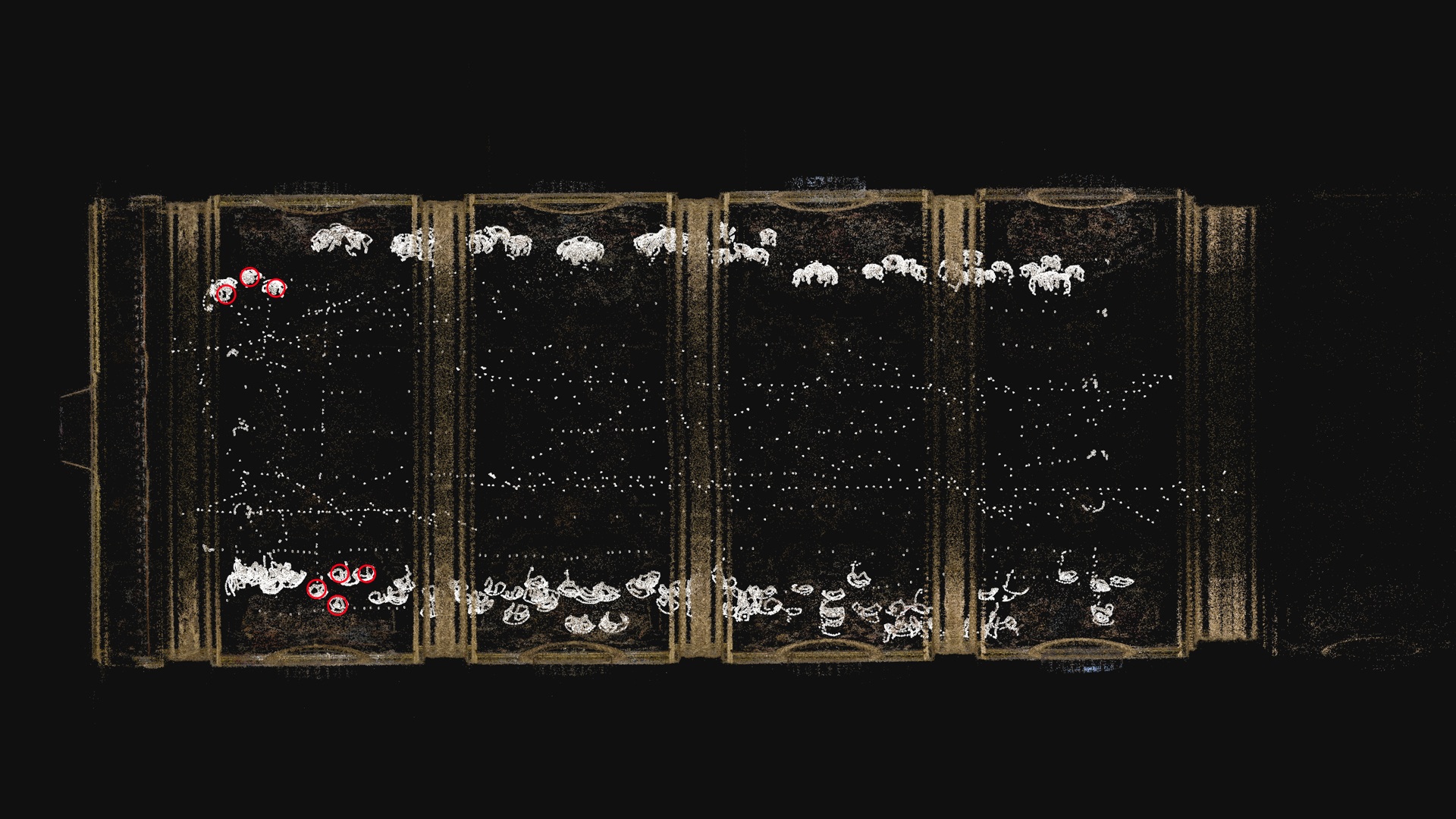}
    \caption{Camera positions and lift stations in the nave}
    \label{fig:site-plan:lifts}
  \end{subfigure}
  \hfill
  \begin{subfigure}[t]{0.48\textwidth}
    \includegraphics[width=\textwidth]{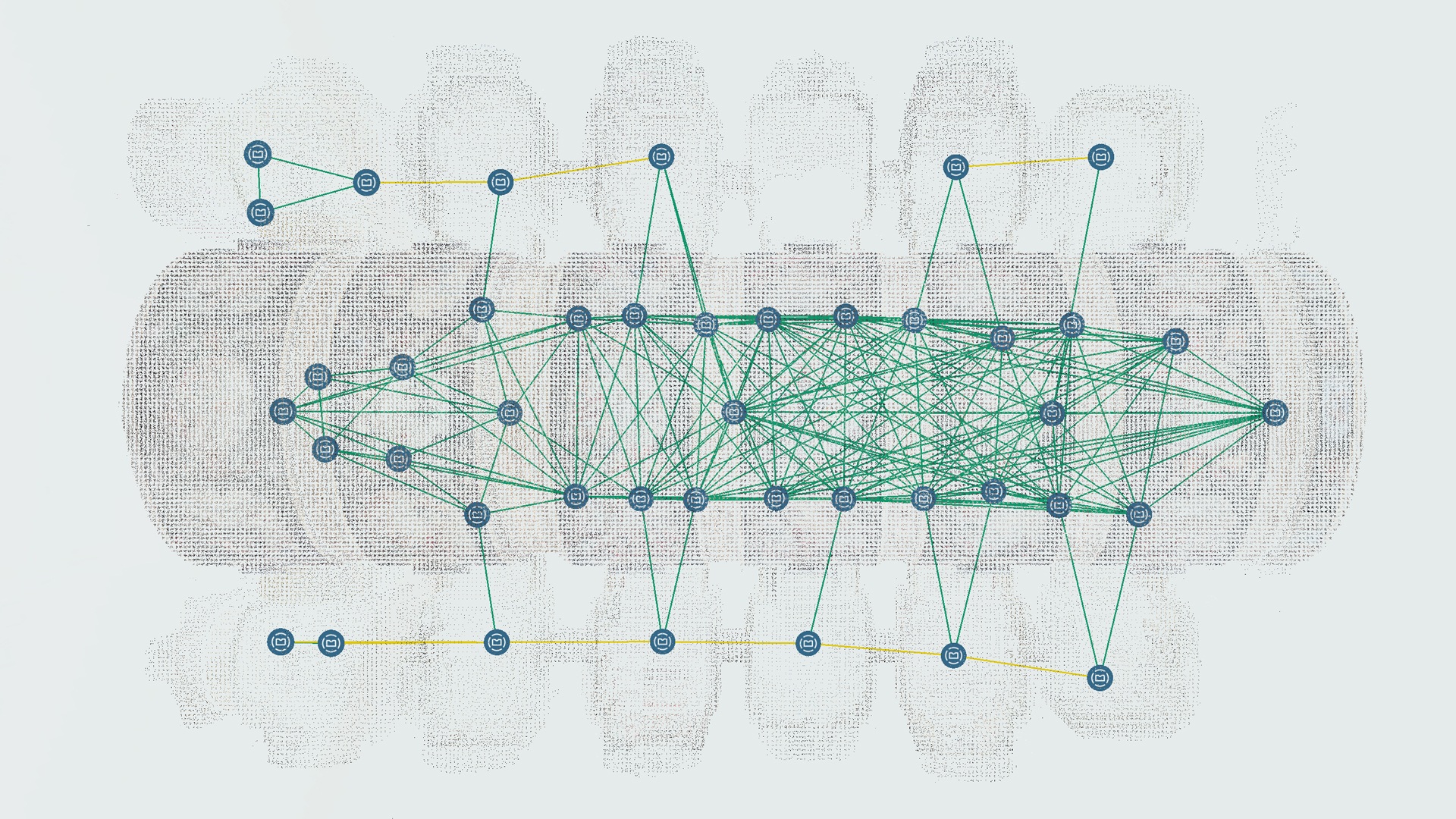}
    \caption{LIDAR scan stations and registration graph}
    \label{fig:site-plan:lidar}
  \end{subfigure}
  \caption{Spatial distribution of data acquisition across St. John's Co-Cathedral. (a) Top-down view from RealityCapture showing aligned camera positions (white points) in the nave. Representative lift stations are circled in red; each cluster of elevated positions corresponds to a lift from which DSLR photographs captured windows, corbels, and capitals. Drone capture was not feasible near the walls due to the risk of disturbing hanging tapestries. (b) LIDAR scan registration from Faro SCENE showing all 43 scan stations with inter-scan alignment connections. Green lines indicate well-registered pairs with high overlap; yellow lines indicate lower-overlap connections at peripheral transitions. The underlying point cloud silhouette reveals the cathedral plan, with a dense mesh of connections across the nave and sparser links to side chapels.}
  \Description{Two-panel figure showing top-down views of camera positions with lift stations circled and LIDAR scan registration network overlaid on the cathedral point cloud}
  \label{fig:site-plan}
\end{figure*}

\begin{figure}[h]
  \centering
  \includegraphics[width=\columnwidth]{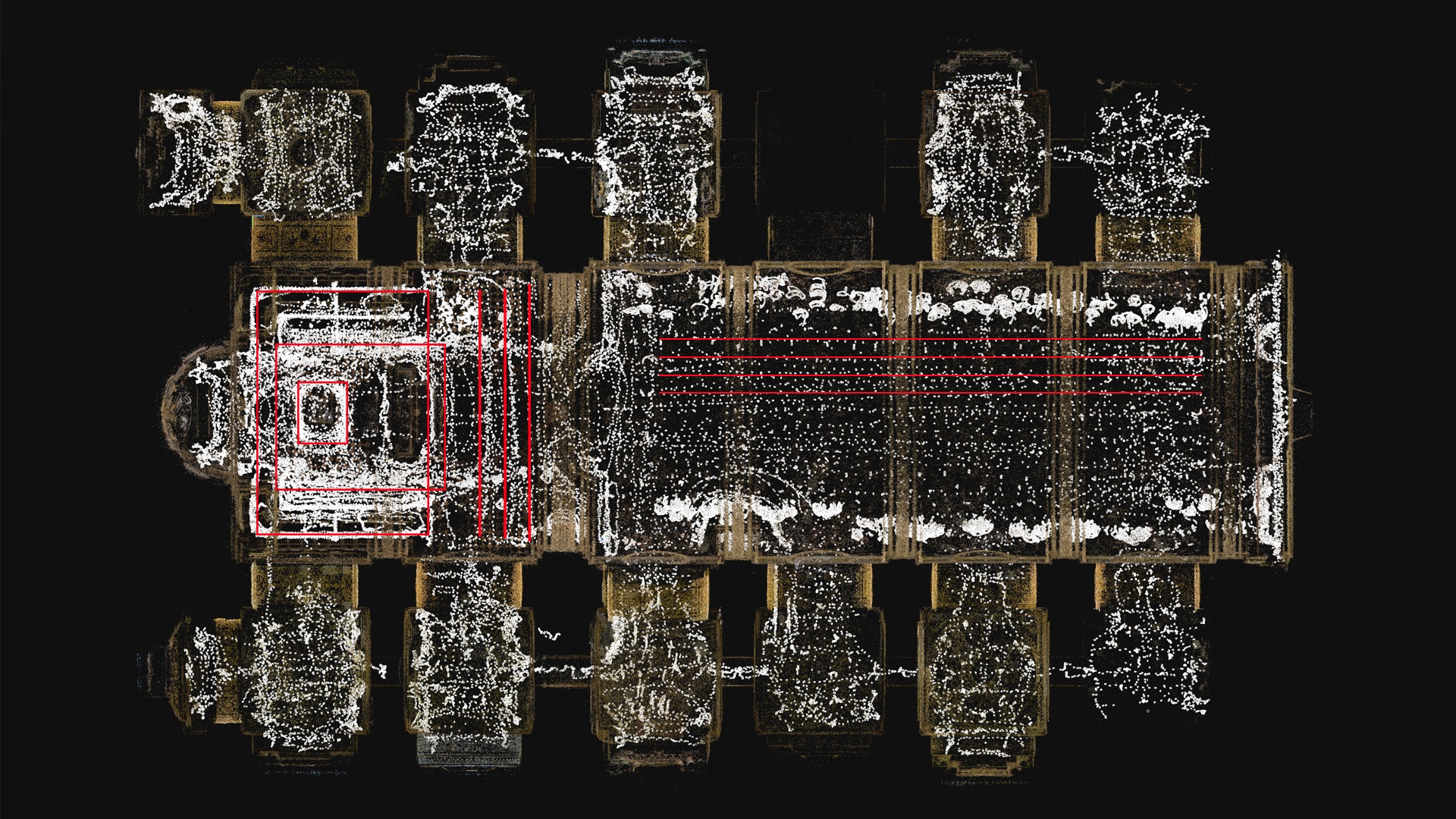}
  \caption{Complete top-down view of all 91,721 aligned camera positions (white points) and drone flight paths (red lines) across the full cathedral, exported from RealityCapture. DSLR positions cluster densely around chapels and along the nave walls. Drone paths trace systematic parallel passes over the nave ceiling and concentric orbits around the altar vault. The full spatial extent of the twelve side chapels, nave, and altar area is visible.}
  \Description{Top-down view of entire cathedral showing all aligned camera positions as white dots with red drone flight path lines overlaid on the point cloud}
  \label{fig:camera-paths}
\end{figure}

Our multi-modal approach (Figure~\ref{fig:data-acquisition}) utilized Canon EOS R5 cameras with 24\,mm lenses at native 8K resolution, including from scaffolding and lifts for ceiling coverage. Figures~\ref{fig:site-plan} and~\ref{fig:camera-paths} show the spatial distribution of all capture positions across the cathedral. A DJI Mavic 3 Pro Cine drone provided elevated perspectives at 5K resolution from altitudes of 3--15\,m, with flight paths planned manually rather than using automated overlap settings; drone use near chapel walls was not possible due to the risk of disturbing hanging tapestries. A Faro Focus M70 LIDAR scanner delivered geometric structure through 60-minute scans from central positions, each producing over 100 million points. Traditional photogrammetry requires circling objects from every direction \cite{Remondino2014}, impossible within St. John's architectural constraints. We developed a modified approach beginning with wide shots to establish context, then moving incrementally closer while maximizing object coverage in early photos to aid pixel matching \cite{Remondino2014}. For chapels, we began shooting from outside entrances to maintain alignment with the nave, despite tapestries partially blocking sight lines.

Practical challenges included moveable obstructions (chairs, signs, barriers) requiring repositioning each session, affecting capture patterns. Low lighting necessitated prioritizing sharp focus, critical for mesh alignment \cite{Nocerino2017}, over low ISO, accepting grain as necessary trade-off. Evening-only access demanded efficient workflow through advance equipment organization and planned shoot sequences.

With 99,000 images from multiple sources, rigorous organization was essential. We structured data day-by-day with separate directories for each device: 1 LIDAR, 2 DSLRs across 55 SD cards, and 1 drone. Data transferred to central server after each session for backup. The dataset scale necessitated a proxy-based workflow. We generated lower-resolution proxies for all DSLR and drone images, enabling faster review and sorting into room directories. A custom script established systematic naming conventions based on capture day, SD card number, and proxy-to-raw filename mapping. We maintained separation by dumping SD cards after completing each room. LIDAR data required minimal organization, as software combined all scans into unified point clouds. Each scan was numbered sequentially and placed in corresponding directories.

\begin{table}[h]
\centering
\caption{Data Acquisition Statistics}
\label{tab:data-stats}
\begin{tabular}{|l|r|}
\hline
\textbf{Metric} & \textbf{Value} \\
\hline
Total Nights & 7 \\
Total Images Captured & $\sim$99,000 \\
Images Imported to RC & 93,066 \\
DSLR Images & $\sim$85,000 \\
Drone Images & $\sim$14,000 \\
LIDAR Scans & 43 \\
SD Cards Used & 55 \\
Registered Images & 91,721 \\
Registration Rate & 98.6\% \\
\hline
\end{tabular}
\end{table}

Table~\ref{tab:equipment-specs} provides detailed equipment specifications and capture parameters. Cathedral lighting varied substantially between areas, preventing consistent ISO settings; we prioritized sharp focus and accepted higher noise, which was addressed through AI denoising in post-processing. All images were scaled down to 4K resolution prior to alignment to manage computational load, with the processed dataset totalling approximately 2.18\,TB.

\begin{table}[h]
\centering
\caption{Equipment Specifications and Capture Parameters}
\label{tab:equipment-specs}
\begin{tabular}{|l|l|r|}
\hline
\textbf{Device} & \textbf{Parameter} & \textbf{Value} \\
\hline
\multirow{6}{*}{DSLR} & Model & Canon EOS R5 \\
 & Lens & 24\,mm \\
 & Native Resolution & 8192$\times$5464 \\
 & ISO Range & 12800--25600 \\
 & Aperture Range & f/8--f/11 \\
 & Shutter Speed & 1/100--1/250\,s \\
\hline
\multirow{6}{*}{Drone} & Model & DJI Mavic 3 Pro Cine \\
 & Lens & 24\,mm \\
 & Native Resolution & 5272$\times$3948 \\
 & ISO Range & 3200--6400 \\
 & Aperture Range & f/5--f/8 \\
 & Flight Altitude & 3--15\,m \\
\hline
\multirow{3}{*}{LIDAR} & Model & Faro Focus M70 \\
 & Scan Duration & 60\,min \\
 & Points per Scan & 100M+ \\
\hline
\multirow{3}{*}{Processing} & Alignment Resolution & 4K (downscaled) \\
 & Processed Image Size & $\sim$22\,MB / image \\
 & Total Dataset & $\sim$2.18\,TB \\
\hline
\end{tabular}
\end{table}

\section{Processing Pipeline}

\begin{figure*}[t]
  \centering
  \includegraphics[width=\textwidth]{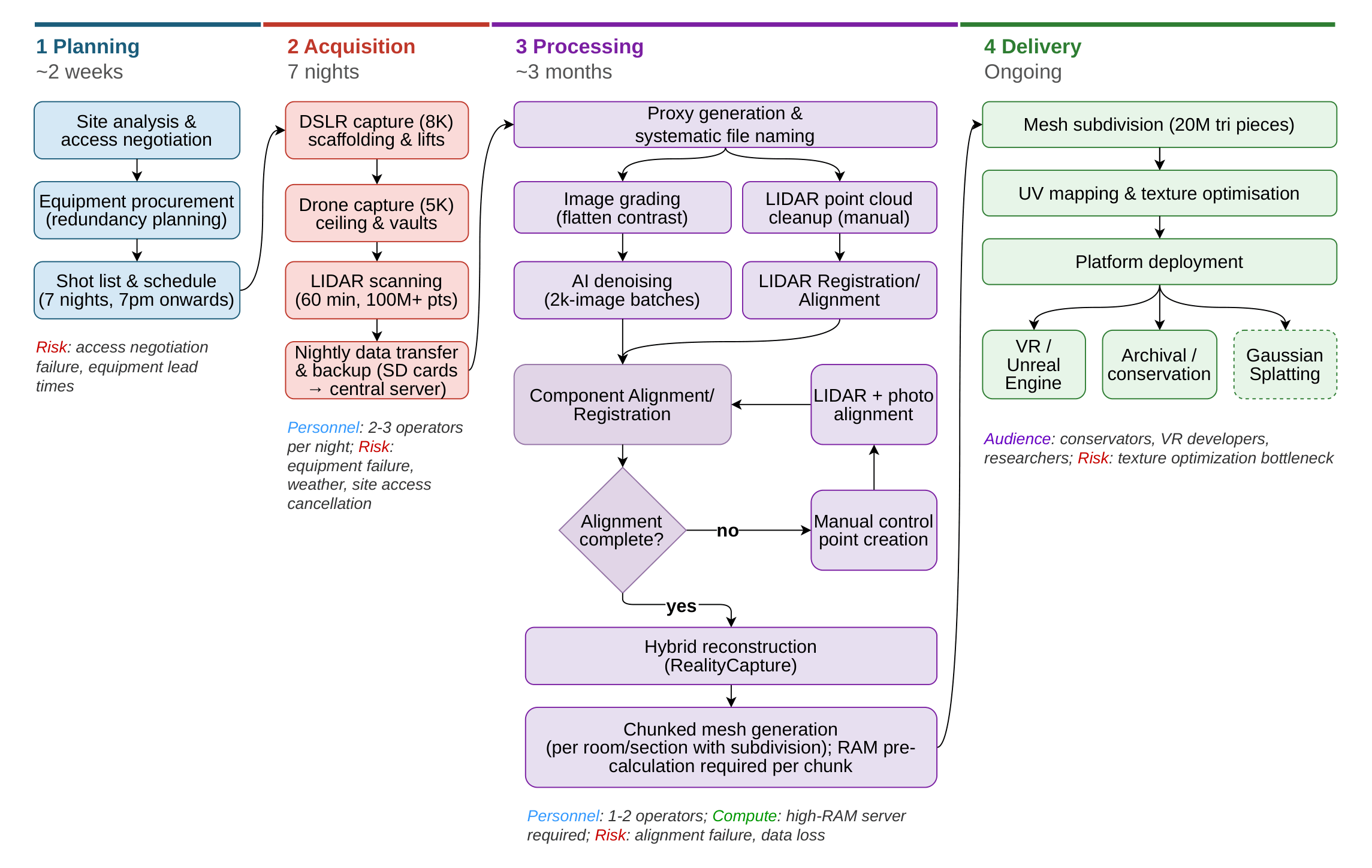}
  \caption{End-to-end workflow diagram for the St. John's Co-Cathedral photogrammetric documentation project. The pipeline spans four phases: (1) Planning and site analysis (2 weeks), including access negotiation, equipment procurement, and shot list preparation; (2) Data acquisition (7 nights), encompassing DSLR, drone, and LIDAR capture with nightly data transfer and backup; (3) Processing (approximately 3 months), including image grading, AI denoising, LIDAR cleanup, hybrid photogrammetric reconstruction, and mesh generation; and (4) Optimization and delivery (ongoing), covering mesh subdivision, texture optimization, and platform-specific deployment. Estimated personnel requirements, computational resources, and key decision points are annotated at each stage to support replicability for heritage professionals planning similar projects.}
  \Description{Workflow diagram showing four project phases from planning through delivery with timing and resource annotations}
  \label{fig:workflow-diagram}
\end{figure*}

Figure~\ref{fig:workflow-diagram} presents our end-to-end workflow, illustrating the complete pipeline from planning through delivery with approximate timelines and resource requirements at each stage. This diagram is intended to support heritage professionals in planning similar projects by identifying key decision points, risk factors, and resource bottlenecks. The processing phase, detailed in this section, constitutes the most time-intensive component.

Proper image grading proved critical for both photogrammetric accuracy and final texture quality. We reduced contrast to flatten the images, minimizing baked-in lighting effects. This serves two purposes: textures respond better to manual lighting in final applications, and photogrammetry algorithms interpret color variation as geometric change. St. John's Cathedral presented particular challenges due to its extremely dark environment. Flattening dark images introduces significant grain, particularly problematic given the predominance of dark materials including bronze, wood, and metal. 

\begin{figure*}[t]
  \centering
  \begin{subfigure}[t]{0.48\textwidth}
    \includegraphics[width=\textwidth]{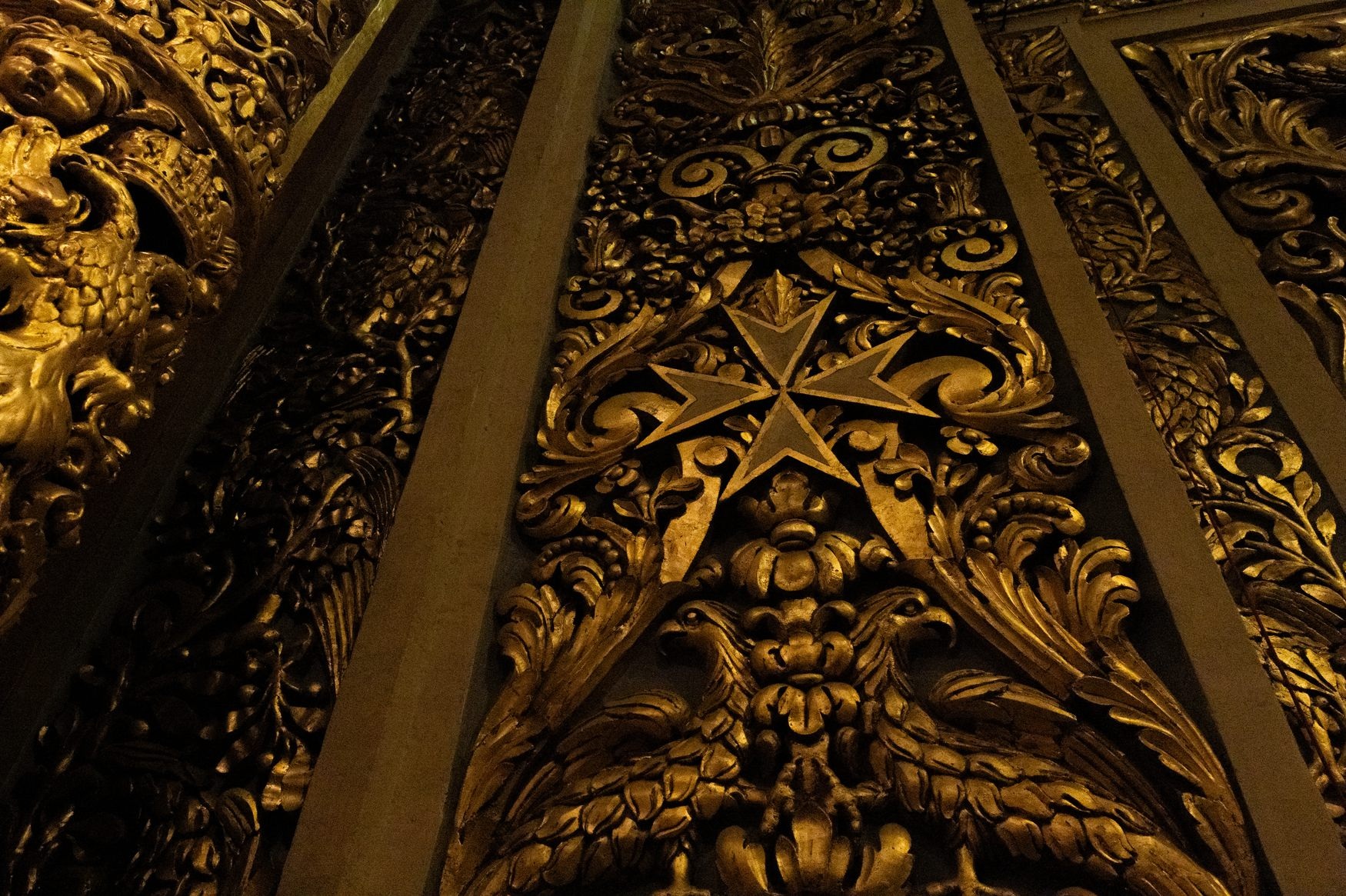}
    \caption{Raw capture with baked lighting}
    \label{fig:reflective:before}
  \end{subfigure}
  \hfill
  \begin{subfigure}[t]{0.48\textwidth}
    \includegraphics[width=\textwidth]{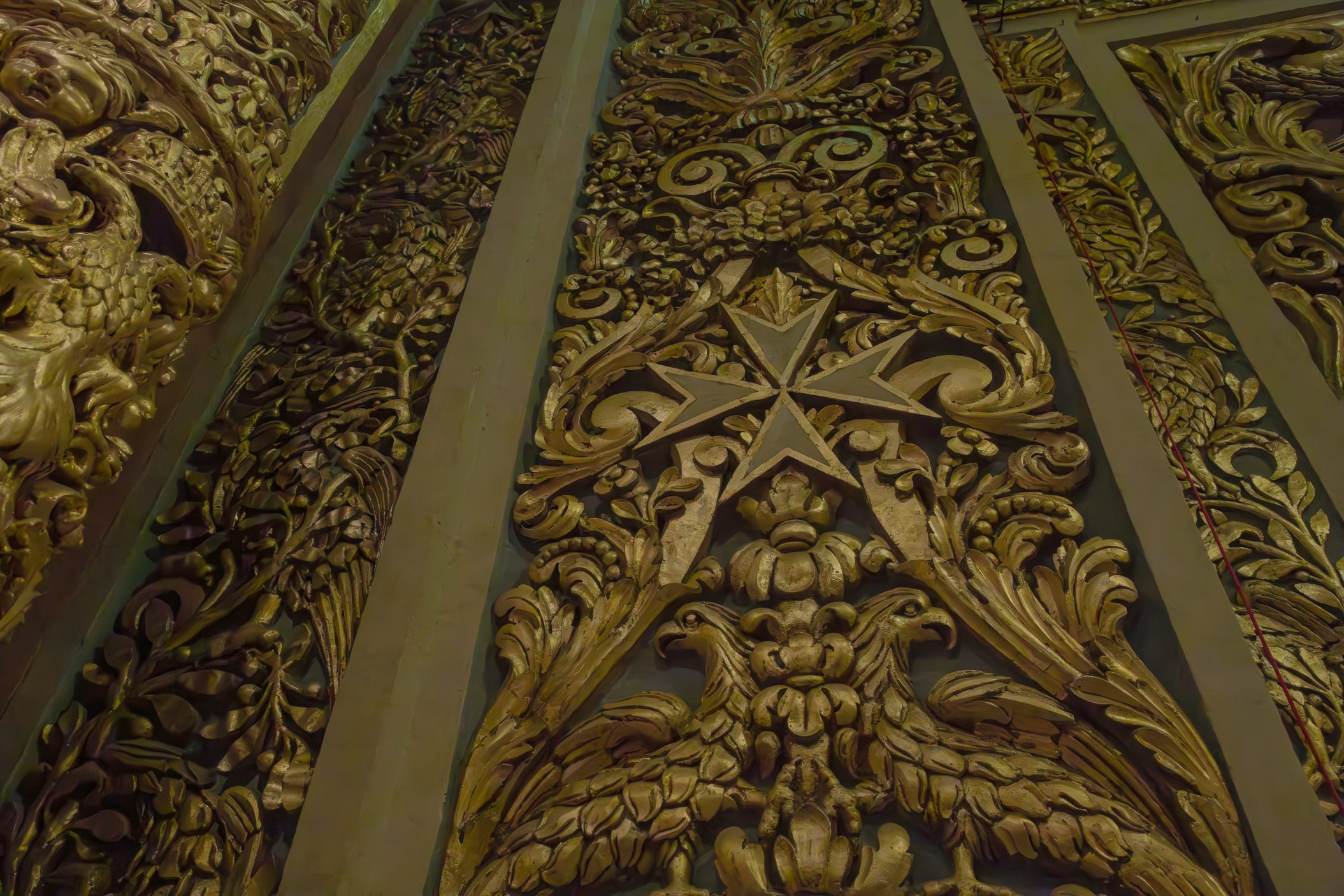}
    \caption{Post-grading with flattened lighting}
    \label{fig:reflective:after}
  \end{subfigure}
  \caption{Reflective surface challenge in ornate gilded decoration featuring the Maltese cross and baroque floral motifs. (a) Raw capture shows strong baked-in lighting with pronounced highlights and shadows on the metallic gold leaf surfaces. (b) After grading, contrast is reduced and lighting is flattened, which improves photogrammetric mesh accuracy and allows for better manual lighting control in final applications. Note the more uniform appearance of the gilt surfaces while maintaining detail definition.}
  \Description{Comparison of ornate gold leaf baroque decoration showing raw capture versus flattened graded version}
  \label{fig:reflective-surfaces}
\end{figure*}

Highly reflective surfaces posed additional challenges, as reflections shift between photos, confusing photogrammetry software which cannot distinguish between actual geometry movement and reflection changes (Figure~\ref{fig:reflective-surfaces}). The raw capture (Figure~\ref{fig:reflective:before}) demonstrates how metallic surfaces create strong specular highlights that change position across different viewpoints, while the graded version (Figure~\ref{fig:reflective:after}) shows reduced contrast that helps minimize these confounding reflections. We established white balance presets for each room and verified accuracy using color charts, matching the color profile of our cameras to maintain consistency across areas. We carefully adjusted highlights and blacks to find common ground across all capture areas.

We employed Adobe's AI denoiser, leveraging recent advances in deep learning-based image denoising \cite{Zhang2017}, to combat grain introduced during grading. The AI denoiser performed excellently on our imagery but proved extremely time-consuming. The software struggled with batches exceeding 2,000 photos, forcing us to process images in 2k chunks. Denoising 99,000 images required approximately one month of continuous processing. 

\begin{figure*}[t]
  \centering
  \begin{subfigure}[t]{0.48\textwidth}
    \includegraphics[width=\textwidth]{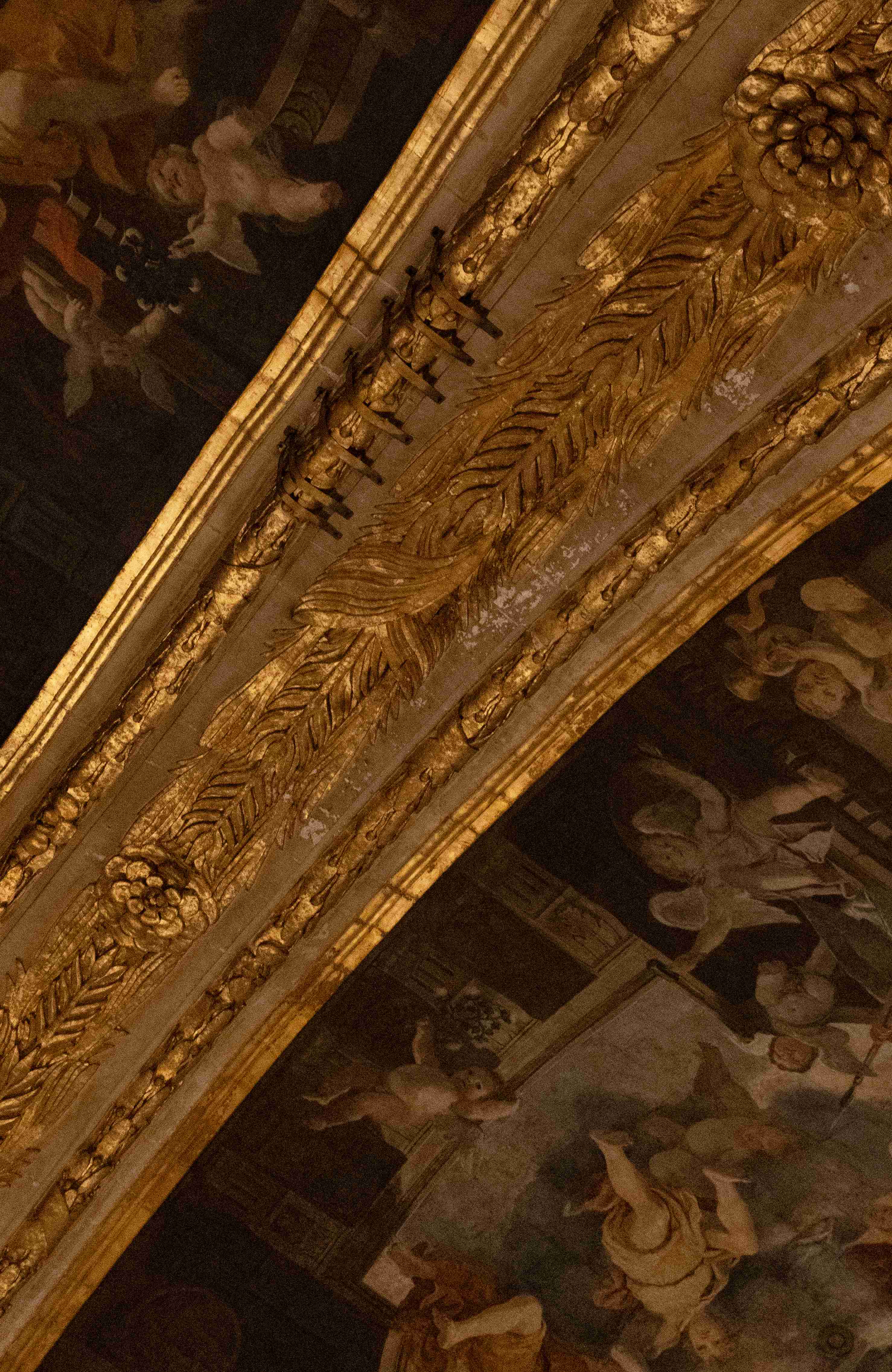}
    \caption{Before denoising - grain visible}
    \label{fig:denoise:before}
  \end{subfigure}
  \hfill
  \begin{subfigure}[t]{0.48\textwidth}
    \includegraphics[width=\textwidth]{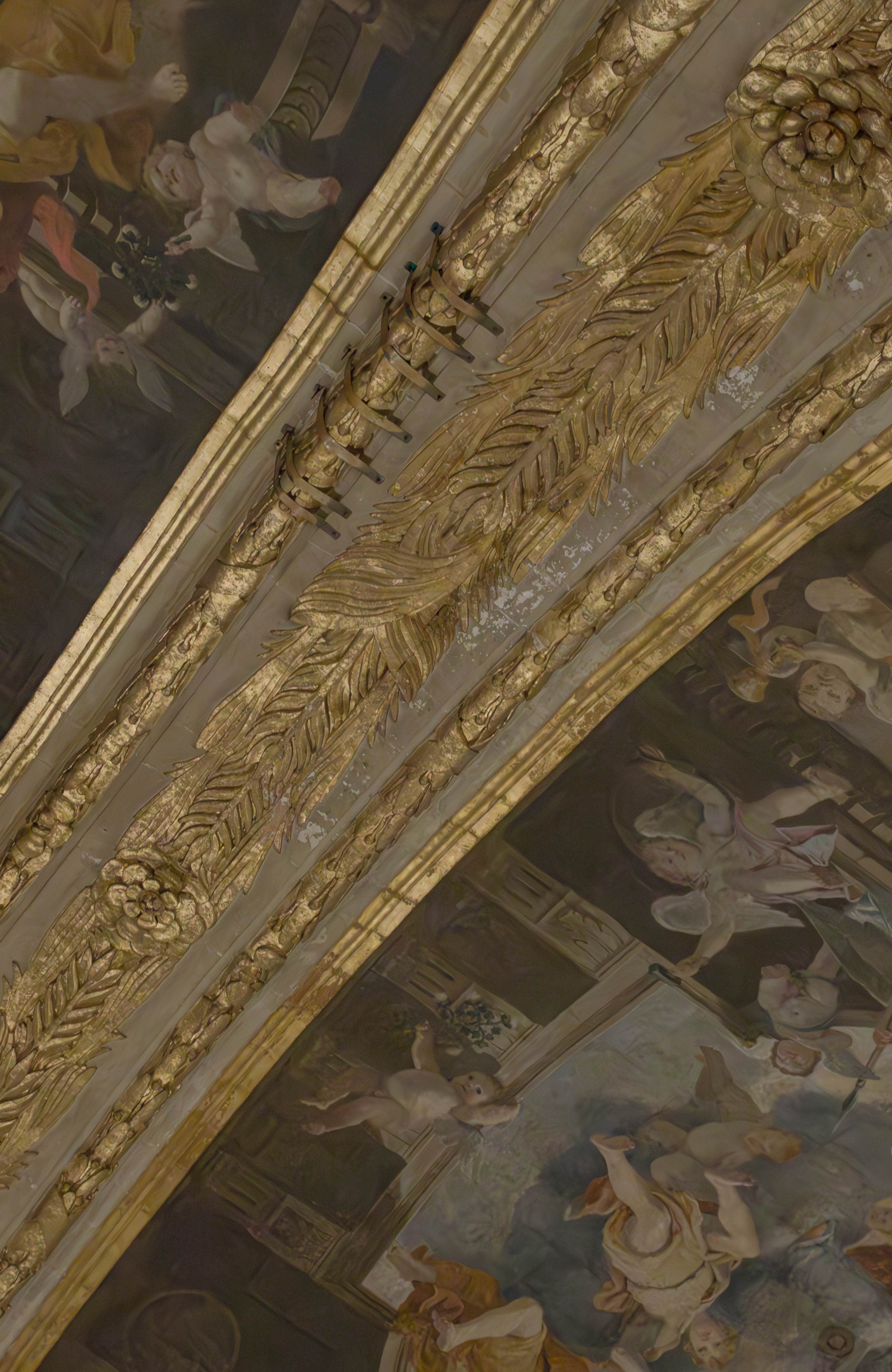}
    \caption{After AI denoising - grain reduced}
    \label{fig:denoise:after}
  \end{subfigure}
  \caption{Image denoising impact on ornate gilded ceiling detail. (a) After image grading to flatten lighting, significant grain is visible throughout the dark baroque decoration, particularly noticeable in the shadow areas and dark painted sections. This grain resulted from the necessary exposure adjustments in the cathedral's extremely low-light environment. (b) After applying Adobe's AI denoiser, grain is substantially reduced while preserving fine architectural details in the gilt ornamental work and painted paintings. The denoising process required one month to process all 99,000 images in 2,000-photo batches.}
  \Description{Comparison of baroque ceiling decoration showing grainy dark image versus clean denoised version}
  \label{fig:denoising}
\end{figure*}

Figure~\ref{fig:denoising} illustrates the dramatic improvement achieved through AI denoising. The pre-denoised image (Figure~\ref{fig:denoise:before}) exhibits substantial grain throughout the darker regions, a direct consequence of flattening the already underexposed imagery from the cathedral's dim interior. After processing (Figure~\ref{fig:denoise:after}), the image maintains fine architectural detail in the gilt work while presenting much cleaner dark areas, producing textures more suitable for both photogrammetric reconstruction and final visualization.

LIDAR data from the Faro Focus M70 required extensive manual cleanup before integration with photogrammetry. Each of the 43 sixty-minute scans (for example, 10 scans for the altar area alone) required individual attention. Reflective surfaces reduced accuracy, creating noise in the point cloud (a characteristic challenge of terrestrial laser scanning in heritage contexts \cite{Lichti2007}). We used the software's accuracy checking and filtering tools to minimize this, but manual removal of unwanted objects proved very time-consuming given the data volume (100+ million points per scan). We removed scaffolding, people, chairs, and other temporary elements. LIDAR scans cast ``shadows'' where objects obstruct the laser, and areas perpendicular to the scanner or far from it contain sparse data. We removed these regions when better scans from alternate angles provided superior coverage. Edges adjacent to ``ghost areas'' (occluded regions) tend to be less accurate and were removed. Reflections sometimes created false points appearing below ground level, requiring removal.

After individual cleanup, all 43 scans were aligned and combined into a master point cloud. We manually selected common faces between scans, analogous to control points in RealityCapture. No common ground existed between chapel scans, complicating alignment. Across the full project, mean point error reached 1.3mm with a maximum of 3.2mm at peripheral scan pairs, while minimum overlap between any two connected scans was 14.3\% at chapel-to-nave transitions. The core nave registration (28 scans) achieved tighter results: 0.9mm mean point error and 26.1\% minimum overlap. Target-based registration between scan clusters yielded a mean distance error of 5.8mm and mean angular error of 0.75\textdegree, with inclinometer mismatches consistently below 0.011\textdegree{} across all scans. For context, photogrammetry best practices recommend 60-80\% overlap between consecutive images; our LIDAR overlap was necessarily lower due to architectural constraints separating scan positions.

We used RealityCapture for photogrammetric processing, integrating LIDAR and photographic data following established hybrid approaches \cite{Remondino2011, Guidi2014}. LIDAR points have inherent contrast (individual discrete points), making them straightforward to integrate into the photogrammetric pipeline, where they provided a structural backbone for aligning photographic components. Photo alignment didn't always succeed automatically, necessitating manual control point creation. We identified unaligned photo components, located the same physical point in photos from different components, and created control points linking these photos. Denoising softens images, making identical details appear different across photos, so control points must be sharp and distinctive. A minimum of 3 points from different images is required; more photos increase accuracy \cite{Chiabrando2016}. We used 3 photos per control point, 3 control points per component group for reliability. Ultimately, 91,721 of 93,066 imported images (98.6\%) were successfully registered into a single component, yielding 137.2 million tie points with a mean reprojection error of 0.26 pixels. A further $\sim$6,000 images from the original 99,000 captured were excluded prior to import during manual quality review. 108 manually placed control points bridged components that failed automatic alignment. Total alignment required approximately 24 hours: 1.5 hours for feature detection and 22.5 hours for registration.

Table~\ref{tab:accuracy-metrics} consolidates accuracy metrics across all capture and processing modalities.

\begin{table*}[t]
\centering
\caption{Project Accuracy Metrics and Tolerances}
\label{tab:accuracy-metrics}
\begin{tabular}{|l|l|r|l|}
\hline
\textbf{Category} & \textbf{Metric} & \textbf{Value} & \textbf{Notes} \\
\hline
\multirow{5}{*}{LIDAR Registration}
 & Total scans & 43 & 60 min each \\
 & Mean point error & 1.3 mm & Full cathedral \\
 & Max point error & 3.2 mm & Peripheral scan pairs \\
 & Minimum scan overlap & 14.3\% & Chapel-nave transitions \\
 & Mean target distance error & 5.8 mm & Between scan clusters \\
\hline
\multirow{5}{*}{Photogrammetry}
 & Images registered & 91,721 / 93,066 & 98.6\% success rate \\
 & Mean reprojection error & 0.26 px & RealityCapture v1.5.1 \\
 & Max reprojection error & 2.00 px & Filtering threshold \\
 & Tie points & 137.2M & Single component \\
 & Control points used & 108 & Manual alignment \\
\hline
\multirow{2}{*}{LIDAR--Photo Registration}
 & Mean angular error & 0.75\textdegree & Between scan clusters \\
 & Inclinometer mismatch & $<$0.011\textdegree & All 43 scans \\
\hline
\multirow{3}{*}{Processing Time}
 & Feature detection & 1h 37m & \\
 & Registration & 22h 38m & \\
 & Total alignment & 1d 0h 16m & Single component \\
\hline
\multirow{2}{*}{Final Reconstruction}
 & Total triangles & 25--30 billion & Full cathedral \\
 & Per-section triangles & $\sim$1 billion & Per room section \\
\hline
\end{tabular}
\end{table*}

The 99k image dataset couldn't be processed as a single unit. Each chapel was processed individually, then chapels were combined (north side, south side). The altar section, containing 33k photos, was further subdivided into approximately 5 chunks. We calculated RAM requirements before processing to prevent crashes and optimize chunk sizes. Point clouds were then divided into regions for mesh generation, creating separate projects per room, with each room containing multiple meshes. For example, the altar was split into 9 pieces using an automated plugin with small overlaps, processed as one project with a mesh per piece. Chapels required manual splitting due to size variations across the 12 distinct spaces. The nave was divided into 5 sections. Each section comprised approximately 1 billion triangles, yielding a total of 25-30 billion triangles across the entire cathedral.

\begin{figure*}[t]
  \centering
  \begin{subfigure}[t]{0.24\textwidth}
    \includegraphics[width=\textwidth]{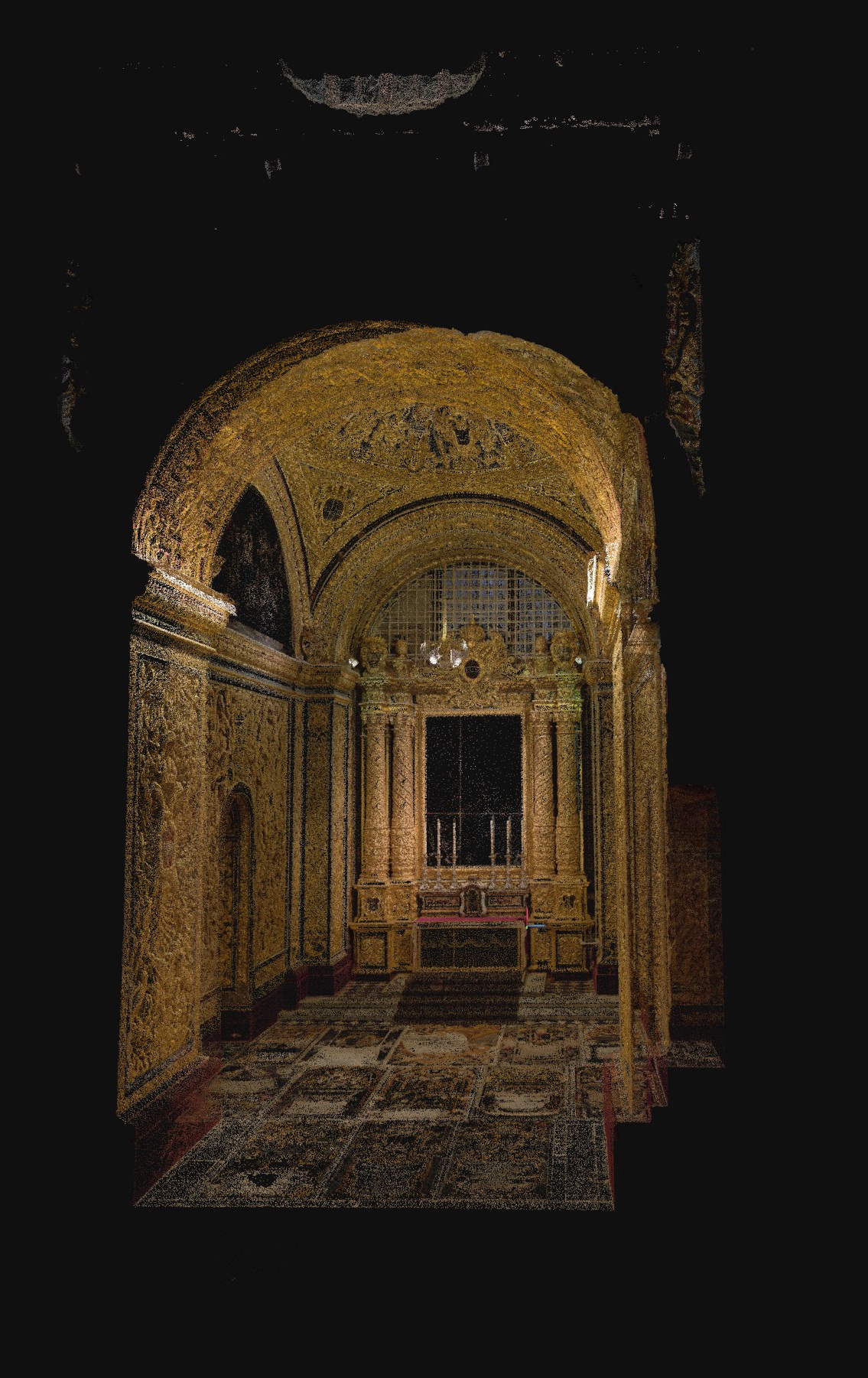}
    \caption{Sparse point cloud with gaps}
    \label{fig:progression:sparse}
  \end{subfigure}
  \hfill
  \begin{subfigure}[t]{0.24\textwidth}
    \includegraphics[width=\textwidth]{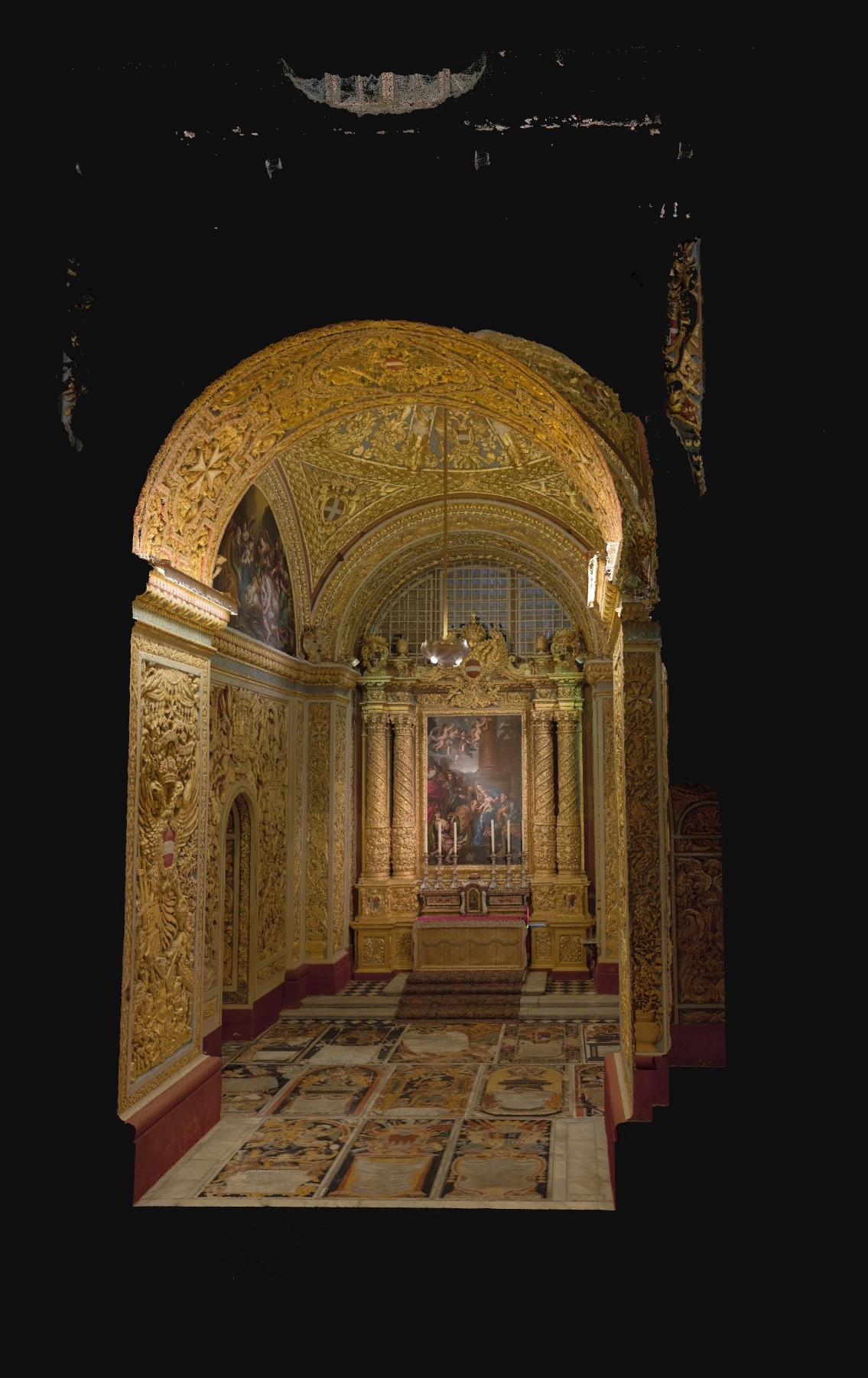}
    \caption{Dense point cloud}
    \label{fig:progression:dense}
  \end{subfigure}
  \hfill
  \begin{subfigure}[t]{0.24\textwidth}
    \includegraphics[width=\textwidth]{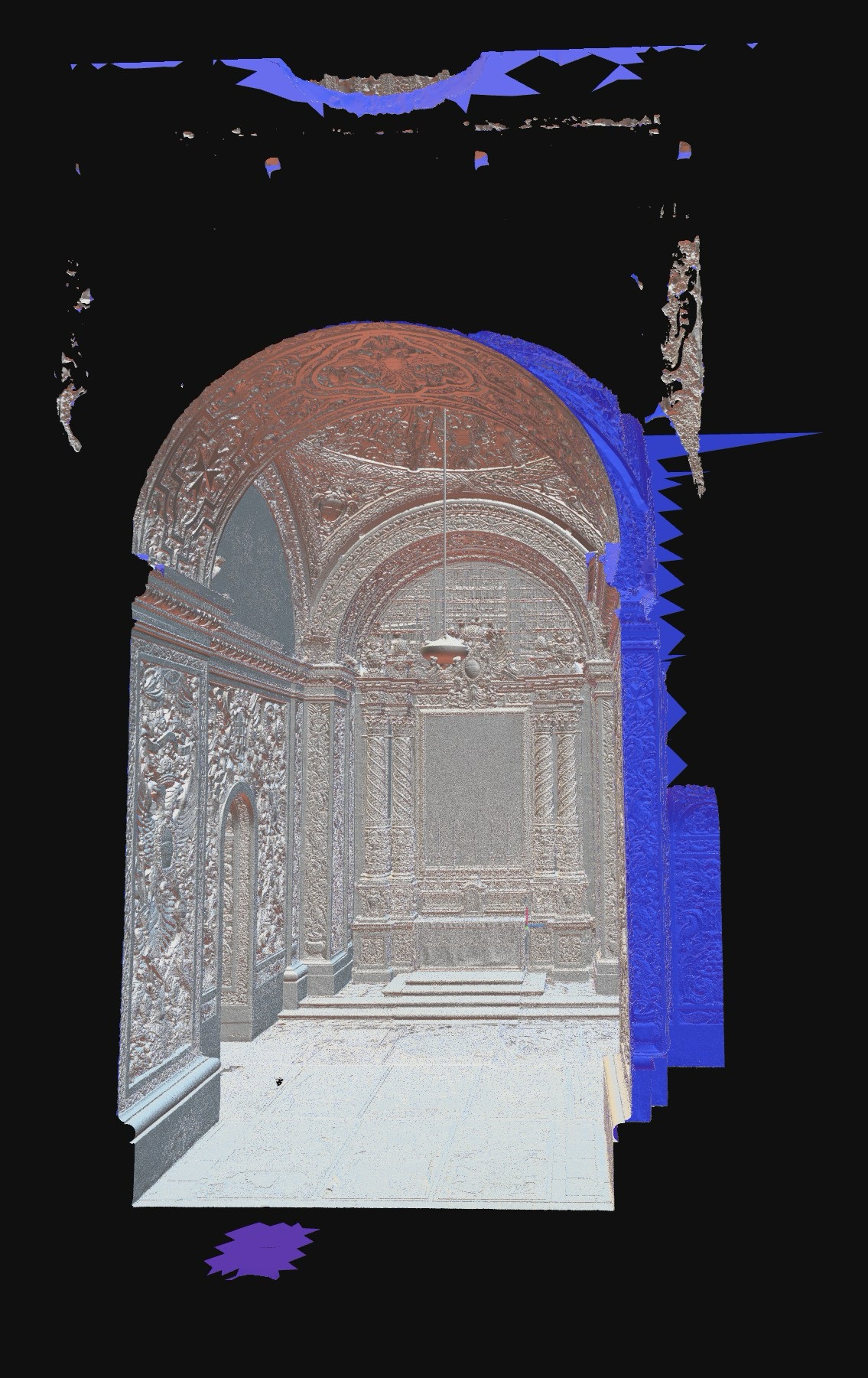}
    \caption{Processed point cloud with alignment}
    \label{fig:progression:aligned}
  \end{subfigure}
  \hfill
  \begin{subfigure}[t]{0.24\textwidth}
    \includegraphics[width=\textwidth]{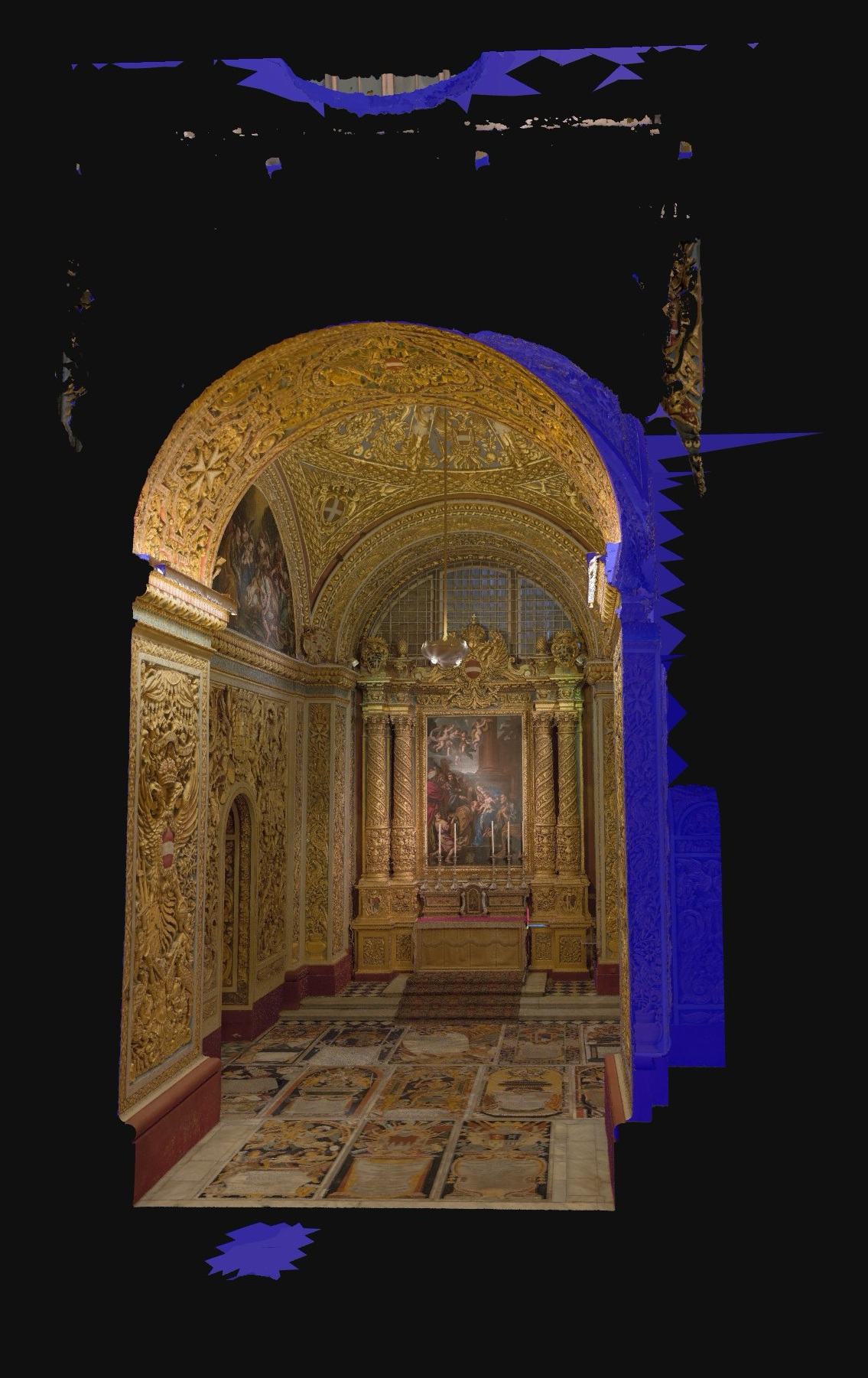}
    \caption{Final textured mesh}
    \label{fig:progression:textured}
  \end{subfigure}
  \caption{Point cloud to mesh reconstruction progression for a chapel section. (a) Initial sparse point cloud shows incomplete coverage with gaps in the reconstruction. (b) After integrating additional photographs and LIDAR data, a denser point cloud emerges with improved surface coverage. (c) Further processing aligns all photographic components and cleans the point cloud, revealing more complete architectural geometry. Blue regions visible here and in (d) represent back faces of the cropped mesh section. (d) Final mesh generation with texture mapping produces a detailed digital model capturing the ornate gilded altar, decorative archway, and intricate marble floor patterns. The complete chapel reconstruction contains approximately 1 billion triangles.}
  \Description{Four-stage progression showing sparse point cloud evolving to dense cloud to aligned geometry to final textured mesh of baroque chapel}
  \label{fig:point-cloud-progression}
\end{figure*}

Figure~\ref{fig:point-cloud-progression} demonstrates the complete reconstruction pipeline for a chapel section. The progression begins with a sparse point cloud (Figure~\ref{fig:progression:sparse}) showing substantial gaps in coverage where initial photographic alignment was incomplete. As additional images are integrated and LIDAR provides structural scaffolding, the point cloud density increases dramatically (Figure~\ref{fig:progression:dense}). Manual control point creation and component alignment produce a cleaner, more complete point cloud (Figure~\ref{fig:progression:aligned}). Blue regions visible in this and the final stage represent back faces of the cropped mesh section. Finally, mesh generation and texture projection yield the finished reconstruction (Figure~\ref{fig:progression:textured}), capturing intricate details of the gilded baroque altar, ornamental archway, and decorative marble floor.

These meshes were subdivided for real-time engines, as Unreal Engine handles many small objects better than few enormous objects. We divided meshes into hundreds or thousands of pieces, each approximately 20 million triangles. Regions weren't square (following architectural features), requiring careful row/column specification to avoid empty subdivisions. Occasionally we split meshes into sub-regions if pieces remained too large. UV maps were generated for each mesh piece, with high resolution requiring hundreds of UV maps. Textures were initially enormous and unusable without optimization, a challenge we continue to address.

\begin{figure*}[t]
  \centering
  \begin{subfigure}[t]{0.48\textwidth}
    \includegraphics[width=\textwidth]{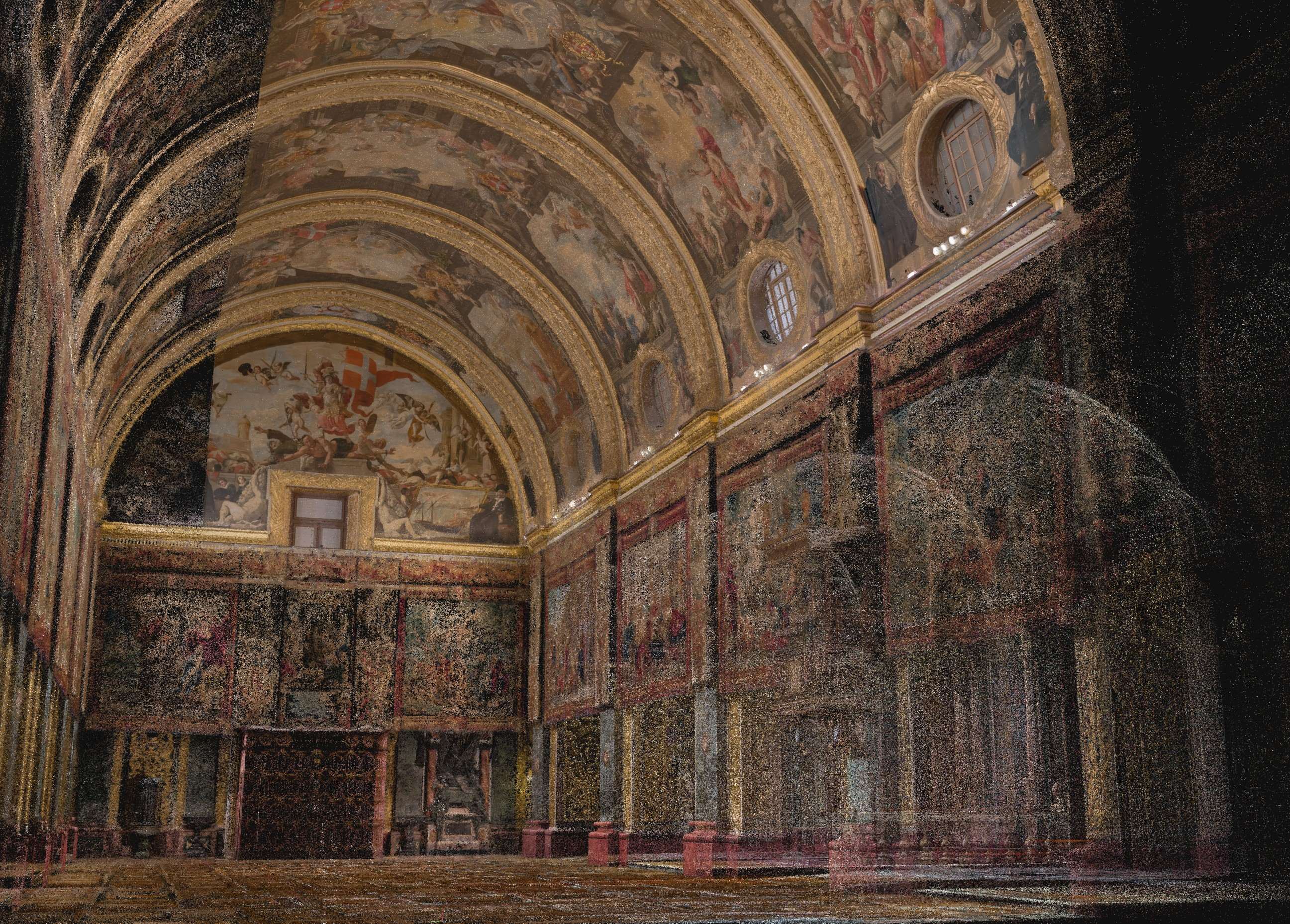}
    \caption{Complete cathedral reconstruction}
    \label{fig:detail:full}
  \end{subfigure}
  \hfill
  \begin{subfigure}[t]{0.48\textwidth}
    \includegraphics[width=\textwidth]{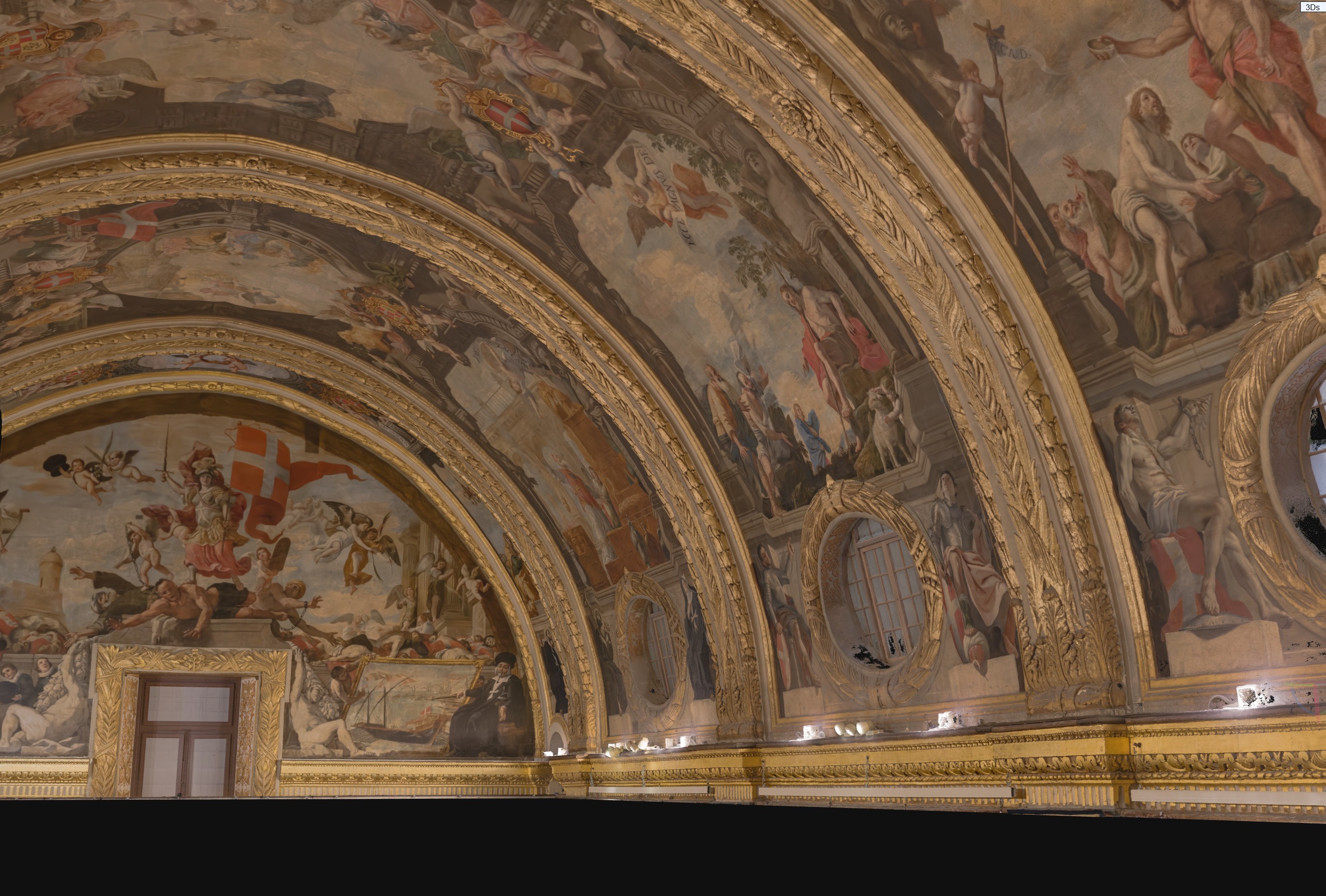}
    \caption{Chapel-level detail}
    \label{fig:detail:chapel}
  \end{subfigure}
  
  \vspace{1em}
  
  \begin{subfigure}[t]{0.48\textwidth}
    \includegraphics[width=\textwidth]{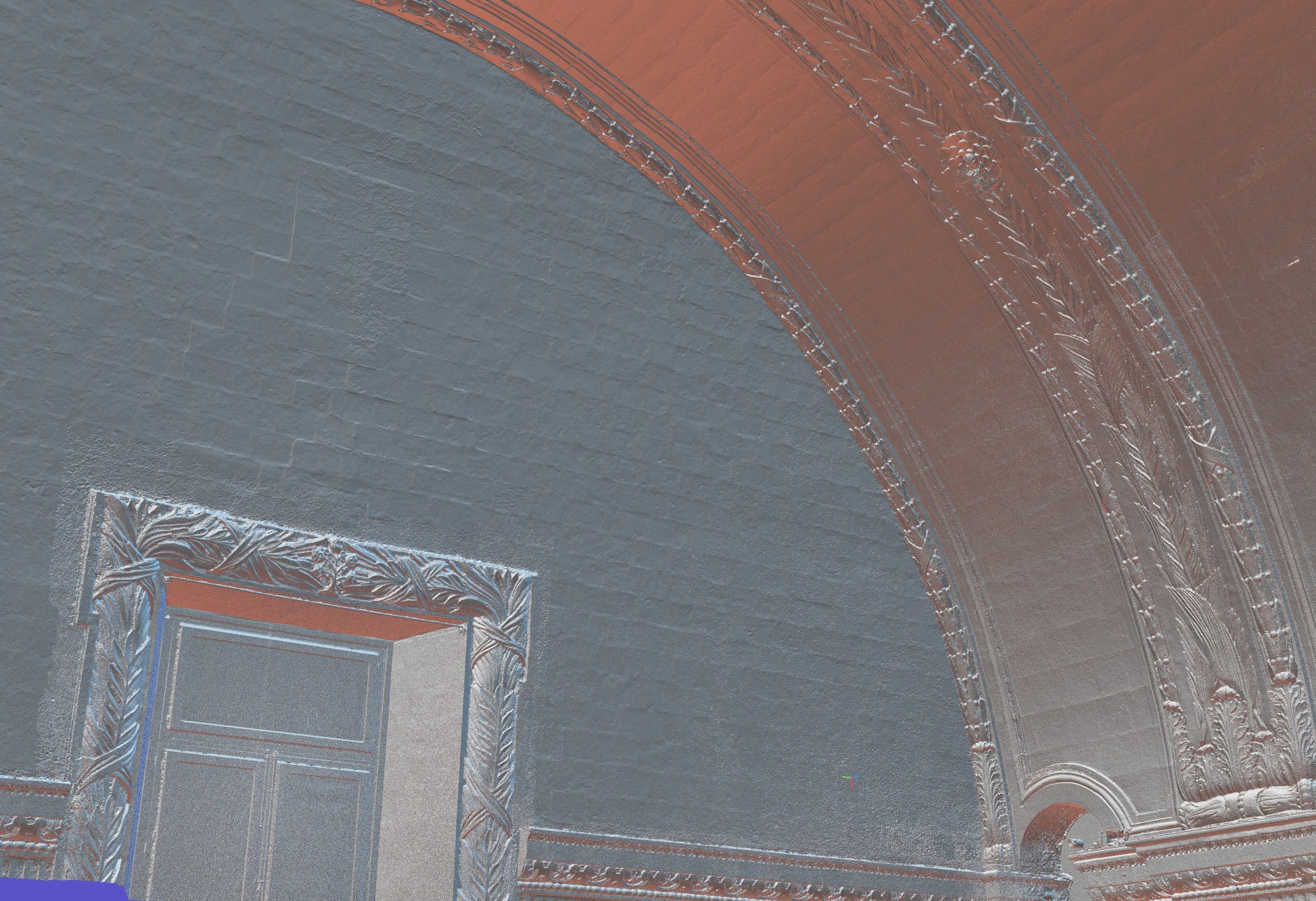}
    \caption{Extreme mesh detail with texture removed}
    \label{fig:detail:mesh}
  \end{subfigure}
  \caption{Multi-scale reconstruction detail from complete cathedral to micro-surface geometry. (a) The complete 25-30 billion triangle reconstruction encompasses all architectural spaces including nave, chapels, and altar areas, demonstrating the scale and completeness of the digital preservation. (b) Chapel-level detail reveals the ornate baroque ceiling vaults, wall tapestries, and intricate gilded architectural elements captured at high fidelity. (c) Extreme closeup with texture removed exposes the underlying mesh geometry, revealing stone surface displacement, cracks, and structural details critical for conservation assessment. This level of geometric detail enables structural integrity analysis and documentation of material degradation for preservation purposes.}
  \Description{Three-panel comparison showing progressive zoom from full cathedral model to chapel detail to extreme mesh closeup revealing stone surface geometry}
  \label{fig:reconstruction-detail}
\end{figure*}

The multi-scale fidelity of our reconstruction (Figure~\ref{fig:reconstruction-detail}) demonstrates the workflow's capability to capture architectural heritage at multiple levels of detail. The complete model (Figure~\ref{fig:detail:full}) preserves spatial relationships across the entire cathedral complex. At the chapel scale (Figure~\ref{fig:detail:chapel}), individual decorative elements, tapestries, and architectural features remain clearly defined. Most significantly, the extreme detail view (Figure~\ref{fig:detail:mesh}) with texture removed reveals the underlying mesh captures micro-surface geometry including stone cracks, surface displacement, and material weathering patterns. This geometric precision enables structural integrity assessment and conservation planning beyond visual documentation alone.

\section{Results and Applications}

Our reconstruction pipeline has produced a comprehensive digital archive deployable across multiple visualization platforms. The 25-30 billion triangle mesh, subdivided into manageable components for real-time engines, enables both archival preservation and interactive exploration.

\begin{figure*}[t]
  \centering
  \begin{subfigure}[t]{0.48\textwidth}
    \includegraphics[width=\textwidth]{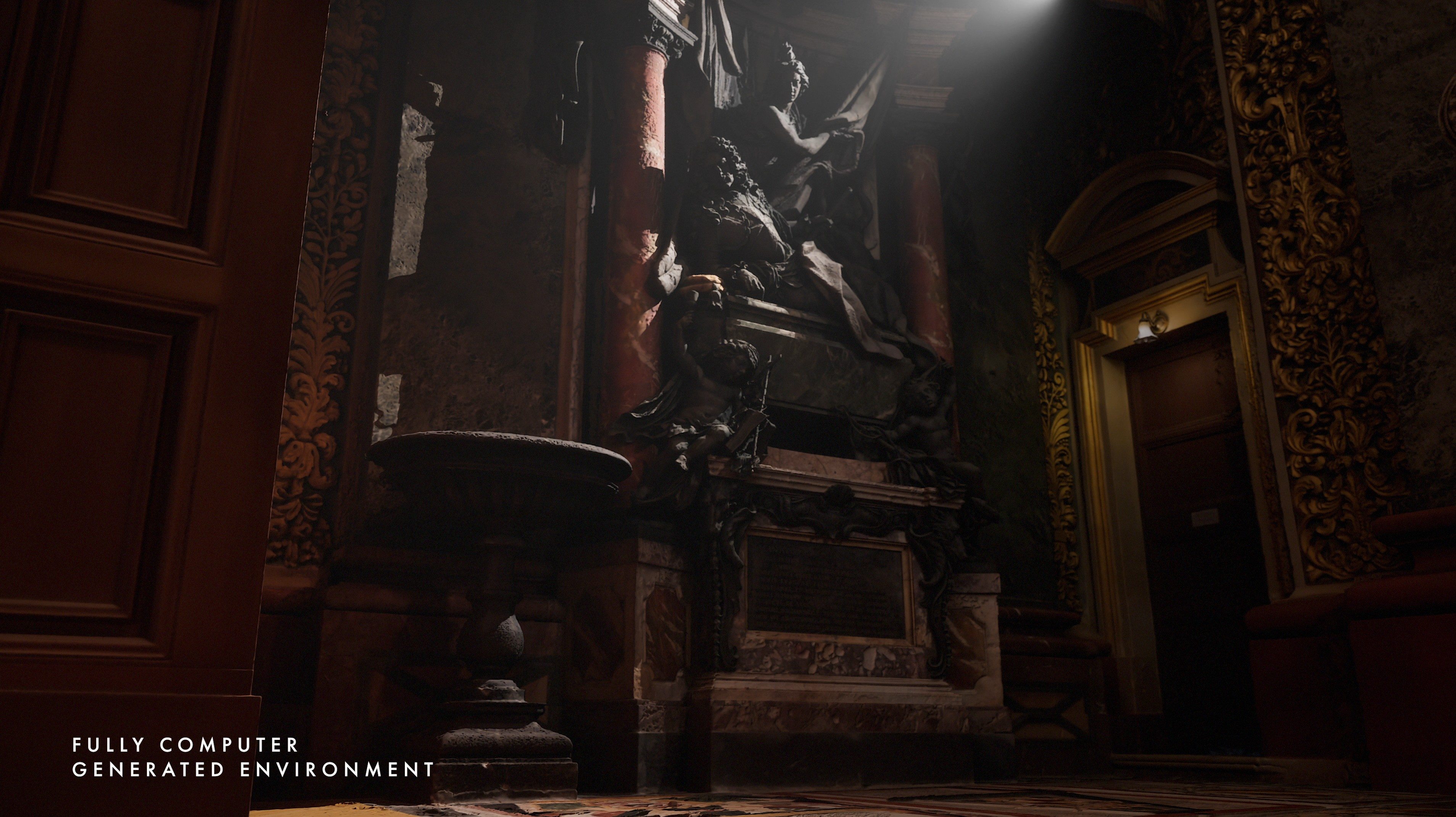}
    \caption{Funerary monument - challenging dark marble}
    \label{fig:results:monument}
  \end{subfigure}
  \hfill
  \begin{subfigure}[t]{0.48\textwidth}
    \includegraphics[width=\textwidth]{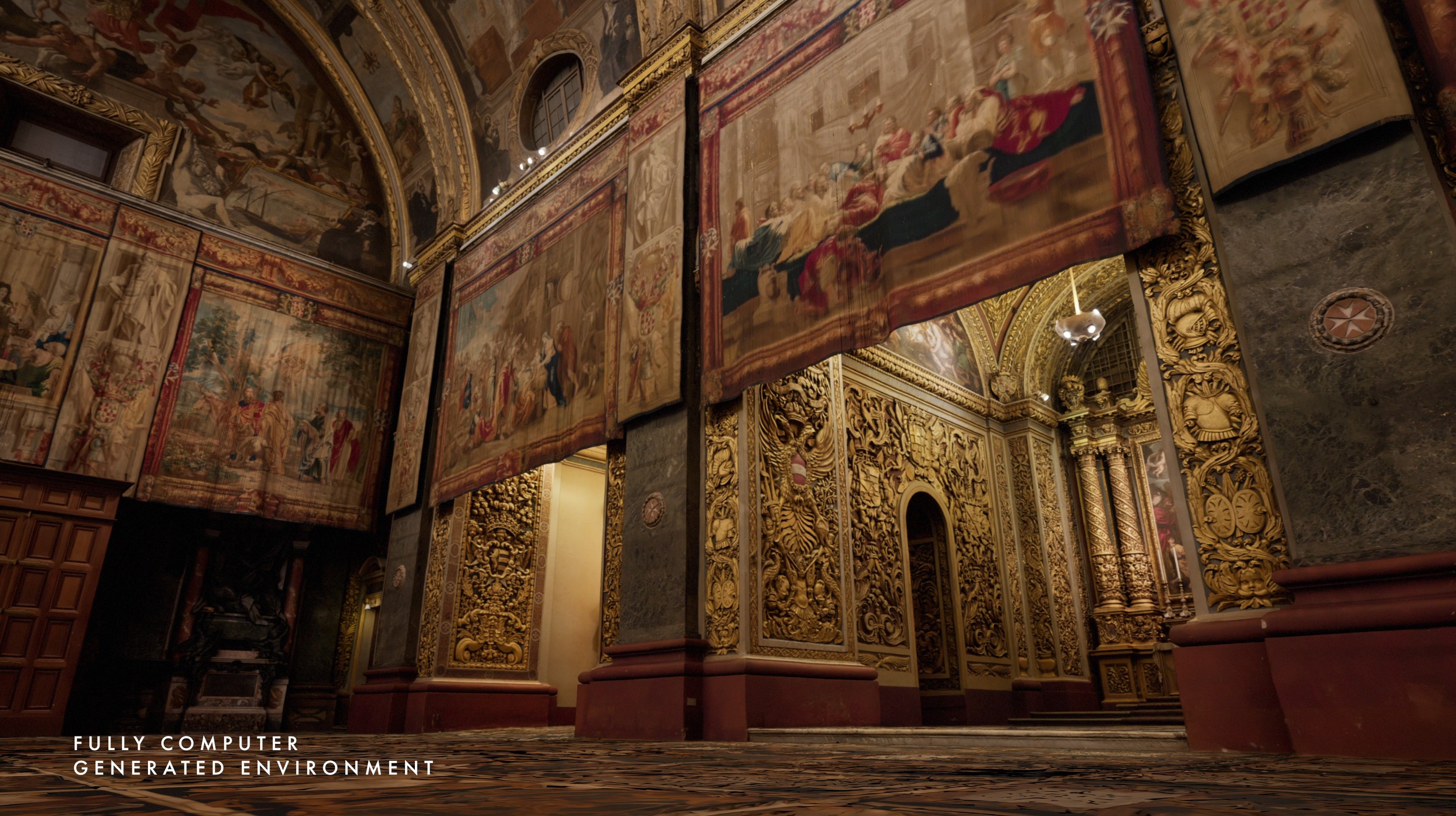}
    \caption{Chapel entrance with tapestries}
    \label{fig:results:chapel}
  \end{subfigure}
  
  \vspace{1em}
  
  \begin{subfigure}[t]{0.48\textwidth}
    \includegraphics[width=\textwidth]{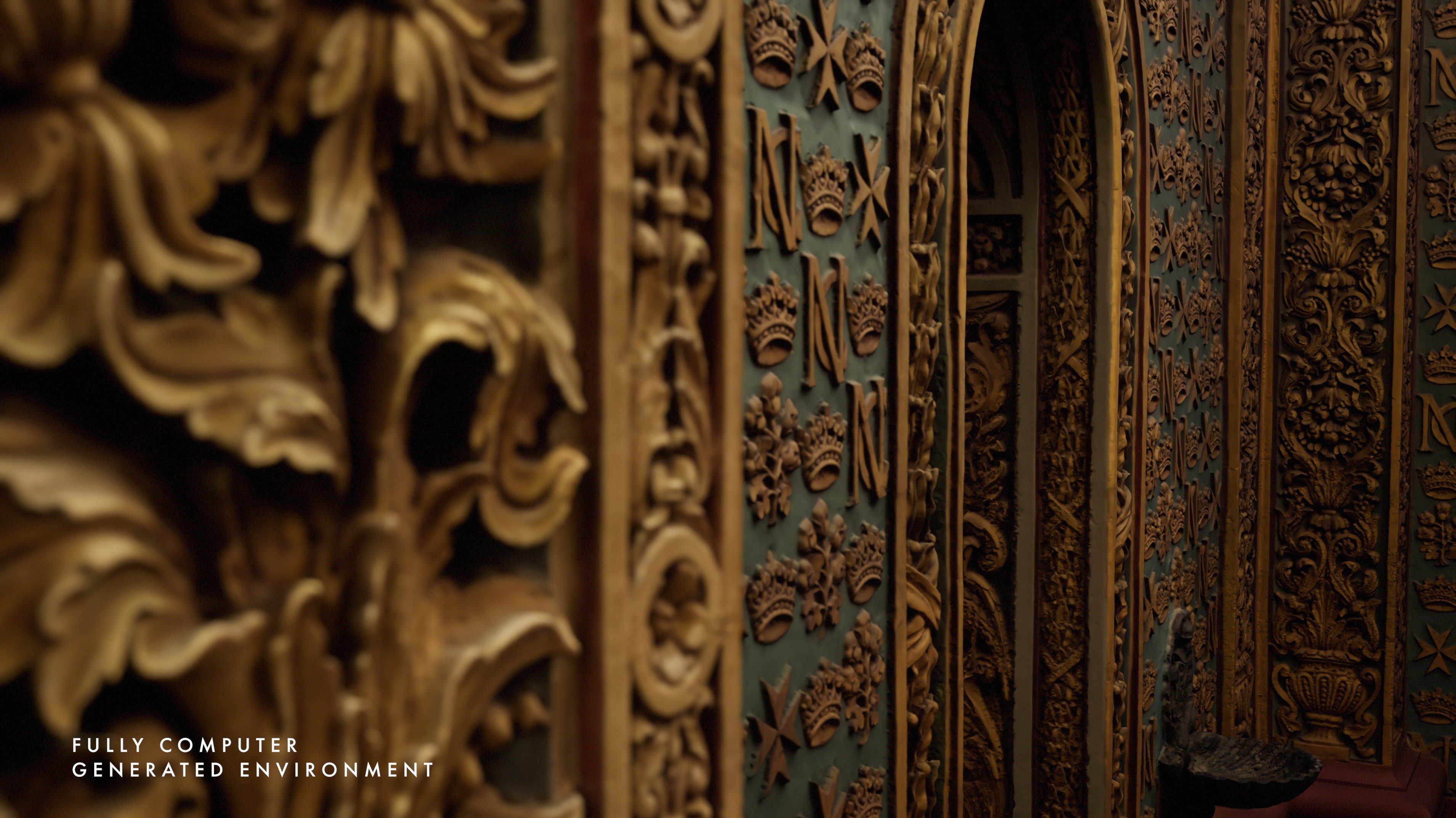}
    \caption{Gilded wall details}
    \label{fig:results:gilded}
  \end{subfigure}
  \caption{Real-time photogrammetric mesh visualization results and remaining challenges. (a) The funerary monument of Grand Master Marc'Antonio Zondadari at the cathedral's main entrance demonstrates challenges with extremely dark, highly reflective marble surfaces. Despite LIDAR support providing geometric structure, imperfections remain visible on curved marble edges, illustrating limitations when processing highly specular materials in low-light conditions. (b) The entrance to the Sacristy and Chapel of Germany showcases reconstructed tapestries. Tapestry quality is compromised due to constant micro-movement of hanging fabric during capture. High-resolution flat photographs of dismantled tapestries will be reprojected onto smoothed meshes in future work. (c) Ornate gilded wall decoration demonstrates high-quality geometric reconstruction of baroque details. While mesh geometry is relatively smooth despite metallic reflections, physically-based rendering (PBR) materials are still required to accurately represent gilt surfaces in interactive applications.}
  \Description{Three rendered views from the photogrammetric mesh reconstruction showing monument with dark marble, chapel entrance with tapestries, and ornate gilded wall decoration}
  \label{fig:processing-results}
\end{figure*}

Figure~\ref{fig:processing-results} illustrates achievements and remaining challenges. The funerary monument (Figure~\ref{fig:results:monument}) presented extreme difficulties: insufficient lighting, highly reflective dark marble, and complex curved geometry pushed our pipeline's limits. While LIDAR provided accurate structure, artifacts remain on curved marble edges where specular reflections confused reconstruction algorithms. The chapel entrance (Figure~\ref{fig:results:chapel}) demonstrates successful spatial capture, but hanging tapestries exhibit quality issues from fabric movement during capture. Our solution involves reprojecting high-resolution photographs of dismantled tapestries onto smoothed meshes. The gilded wall detail (Figure~\ref{fig:results:gilded}) showcases intricate baroque ornamental work captured with excellent geometric fidelity despite reflective gold leaf surfaces, though physically-based rendering materials are required for accurate metallic representation in interactive applications.

\subsection{Gaussian Splatting}

\begin{figure}[h]
  \centering
  \includegraphics[width=\columnwidth]{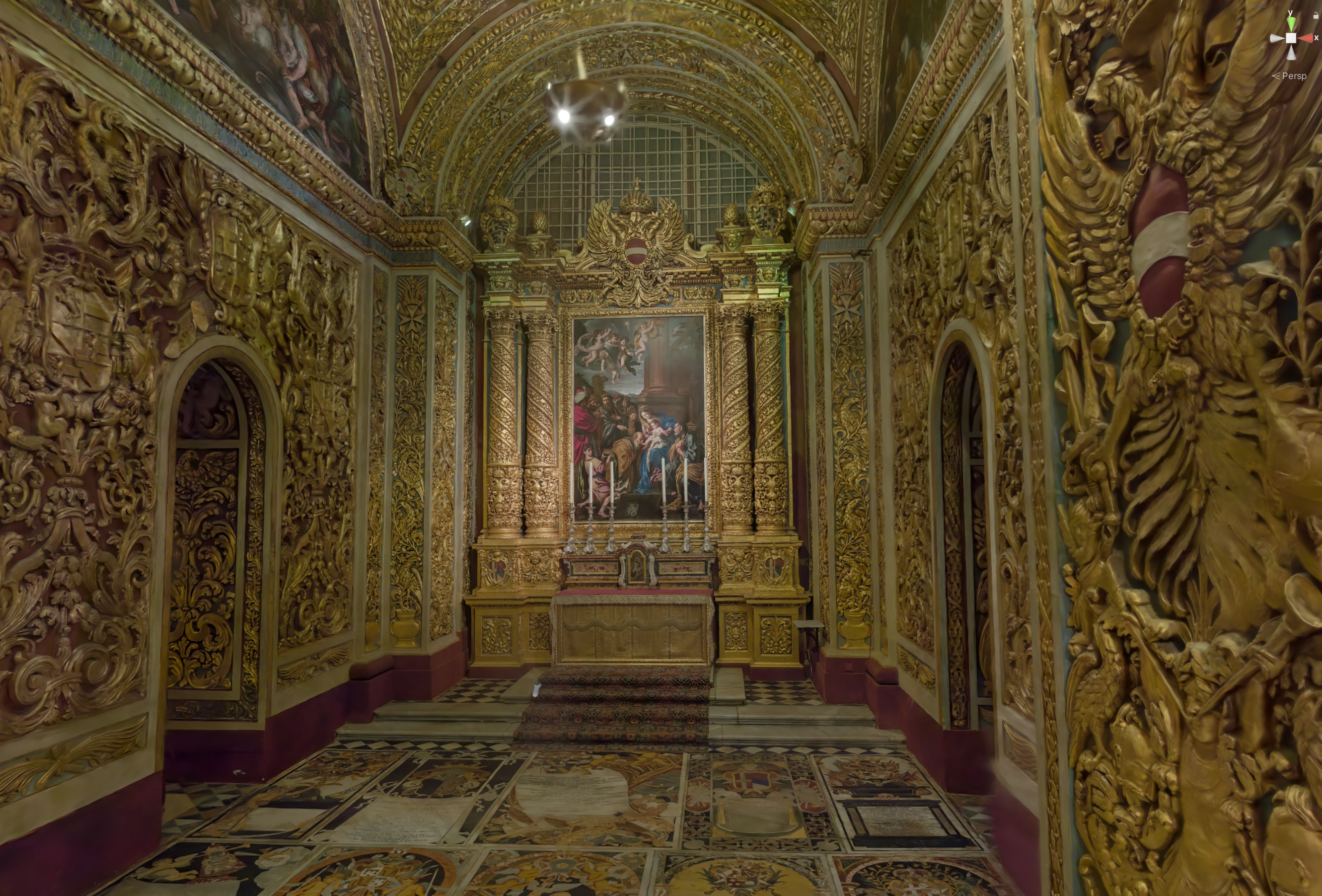}
  \caption{Gaussian splatting reconstruction of the Chapel of Germany. This alternative representation technique demonstrates physically accurate appearance with natural preservation of reflections and specular characteristics on gilt surfaces and tapestries. Gaussian splats excel at capturing the visual complexity of reflective baroque materials that challenged traditional photogrammetric meshing. However, when viewing at close range or moving the virtual camera near surfaces, individual splats become visible and the illusion of solid geometry breaks down. This suggests complementary use cases: Gaussian splats for realistic distant viewing and reflective surface representation, traditional meshes for close inspection and structural analysis.}
  \Description{Gaussian splat rendering of ornate baroque chapel interior showing natural reflections on gold leaf surfaces and tapestries with physically accurate lighting}
  \label{fig:gaussian-splat}
\end{figure}

We conducted preliminary experiments with Gaussian splatting \cite{Kerbl2023} as an alternative representation (Figure~\ref{fig:gaussian-splat}). The Chapel of Germany reconstruction demonstrates strength in preserving physically accurate appearance, particularly reflective and specular characteristics challenging for traditional photogrammetry. Natural reflection preservation on gilt surfaces and complex textile rendering showcase advantages over mesh-based approaches. However, splat representation exhibits limitations at close range, where individual splats become visible. This suggests a hybrid approach: traditional meshes for structural analysis and close inspection, Gaussian splats for realistic distant viewing and reflective surfaces. Future work will explore optimal integration strategies.

\section{Optimization and Future Applications}

Several optimization and application development tasks remain. Textures require significant optimization for real-time applications. For VR applications, meshes must be simplified while preserving visual fidelity, and lighting approaches must balance archival accuracy with practical visualization needs. Multiple texture versions will accommodate various platforms.

The reconstruction serves multiple preservation purposes aligned with international heritage documentation standards \cite{ICOMOS2017, Letellier2007}: disaster recovery documentation for restoration following potential damage (as demonstrated by Notre-Dame \cite{neroulidis2024digital}), conservation planning through detailed structural analysis without physical access, virtual tourism enabling global access, and scholarly research allowing unprecedented examination of architectural and artistic elements. Our methodology specifically addressed challenges unique to baroque heritage sites, particularly complex fabric surfaces and highly reflective metallic elements, providing a replicable workflow for similar projects.

\section{Discussion}

Evening-only access required meticulous planning throughout the project. We recommend pre-visit site analysis to plan equipment placement, detailed shot lists organized by night, backup equipment for every critical component, and buffer time for unexpected obstacles. Regular updates and clear documentation of progress helped maintain alignment with client priorities such as the tapestry detail emphasis. We prioritized sharp focus over low ISO, accepting grain as necessary for accurate mesh alignment. In retrospect, this was correct---mesh accuracy depends on focus, while grain can be addressed in post-processing. Investing time in proxy generation and organization scripts paid dividends throughout processing. This approach is essential for projects exceeding 10k images. Breaking the project into manageable chunks (per room, per section) enabled progress tracking and prevented catastrophic failures.

While automation is valuable, large-scale heritage projects still require significant manual intervention including image sorting and organization, control point creation, LIDAR cleanup, and quality assessment. We identified several areas where better tools would benefit future projects: drone path planning based on preliminary models, automated detection and removal of temporary objects in point clouds, and more robust handling of reflective surfaces in photogrammetry software. Our workflow scales to other large heritage sites with appropriate adjustments and is applicable to other baroque churches, palaces, and archaeological sites with restricted access. However, considerations include computational requirements that increase non-linearly with site size, significant manual cleanup time requiring appropriate budgeting, and access restrictions that may necessitate different capture schedules.

The reconstruction captures the cathedral at a single moment in time; ongoing degradation is not tracked without repeated scanning campaigns. While geometry and color are captured, physical material properties including hardness and composition are not inherent in the model. Temporary elements such as furniture and liturgical items required removal or repositioning, meaning the model doesn't reflect the complete ``in-use'' state. These limitations suggest directions for future heritage digitization projects: incorporating material property capture through multispectral imaging or other sensing modalities, establishing protocols for periodic rescanning to track conservation needs over time, and developing workflows to separately document moveable heritage elements alongside fixed architecture.

\section{Conclusion}

We have presented a comprehensive methodology for large-scale photogrammetric documentation of complex heritage sites, validated through the successful digitization of St. John's Co-Cathedral. Our workflow demonstrates that combining LIDAR scanning with extensive photogrammetry can capture baroque architectural spaces at extraordinary detail (25+ billion triangles) despite significant technical challenges. The documented workflow is scalable to other large heritage sites, provides practical solutions for reflective surfaces and access constraints, accommodates data management strategies for 99k+ image projects, and presents a structured approach to processing and optimization.

This project exemplifies how computer graphics techniques can serve cultural preservation, creating digital archives that will outlive our generation. As heritage sites face threats from climate change, conflict, and degradation, such documentation becomes increasingly critical. The resulting reconstruction enables multiple applications from VR experiences to conservation analysis, with emerging techniques like Gaussian splatting offering promising directions for enhanced visualization. We hope this documented methodology will assist other institutions and researchers in preserving our shared cultural heritage.

\begin{acks}
We thank the St. John's Co-Cathedral Foundation for access and support during this project. This work was also partly funded by Melita Foundation Malta.
\end{acks}

\bibliographystyle{ACM-Reference-Format}
\bibliography{references}








\end{document}